\documentclass{aa}  
\DeclareUnicodeCharacter{2212}{\textminus}
\usepackage{graphicx}
\usepackage{txfonts}
\usepackage{xcolor}

\usepackage{soul}
\usepackage[normalem]{ulem}

\begin{document} 

 \titlerunning{New migration patterns in low-viscosity disks} 
 \authorrunning{M. Sanchez et al.}
   \title{New migration patterns in high planet-star mass ratio systems in disks with low viscosity}
  
   \subtitle{}

    \author{M. Sanchez,
          \inst{1}
          S. Paardekooper,\inst{2}
          N. van der Marel,\inst{1}
           P. Ben\'itez-Llambay,\inst{3}
          G. D. Mulders,
          \inst{4}
          }

   \institute{Leiden Observatory, Leiden University, P.O. Box 9513, 2300 RA Leiden,
The Netherlands\\
           \email{msanchez@strw.leidenuniv.nl}
         \and
             Delft University of Technology, Postbus 5,
2600 AA Delft,
The Netherlands
         \and
        Facultad de Ingenier\'ia y Ciencias, Universidad Adolfo Ib\'añez, Av. Diagonal las Torres 2640, Peñalol\'en, Chile
        \and
        Instituto de Astrof\'isica, Pontificia Universidad Cat\'olica de Chile, Av. Vicu\~na Mackenna 4860, 7820436 Macul, Santiago, Chile
}

   \date{}

 
  \abstract
   {Migration of giant planets remains a complex and rich topic. While significant progress has been made in understanding migration in disks with high viscosity, the migration of planets with high planet-star mass ratios in low-viscosity environments is still not fully understood.}
   {Our aim is to study the migration of planets with high planet-to-star mass ratios embedded in low-viscosity protoplanetary disks, characterized by the viscous parameter $\alpha=10^{-4}$, and to derive analytical prescriptions that could be used to study planet formation across a range of stellar masses, spanning from Sun-like stars to M dwarfs.}
   {We perform hydrodynamical simulations using the \textsc{FARGO3D} code, exploring the migration of planets with high planet-star mass ratios ($10^{-3} \leq q \leq 2 \times 10^{-2}$) under different disk conditions, including variations in gas surface density, scale height, and density slope.}
   {Our simulations reveal a change in the migration direction at a mass ratio of $q \approx 0.002$, with planets exhibiting outward migration for $q > 0.002$. Additionally, for planets undergoing outward migration, we find that the migration speed depends on the unperturbed local gas density. In most cases, outward migration is driven by a positive torque related to the fact that the planet's eccentricity remains below $e < 0.2$. However, under certain disk parameterizations, planets with $q > 0.01$ can develop higher eccentricities in the range $0.2 < e < 0.45$, which can lead to stalled migration.}
   {Our findings suggest that outward migration is a viable mechanism for massive planets in low-viscosity disks, which has implications for the formation and distribution of super-Jupiter planets around Sun-like stars and planets more massive than Neptune around very low-mass stars. Given the challenges in detecting such planets, improving our theoretical understanding of their migration is essential for interpreting exoplanet demographics and guiding future observational efforts.}


   \maketitle

%

\section{Introduction}

Planetary migration, driven by the interactions between planetary embryos and the protoplanetary disk in which they are embedded, is thought to play a crucial role in shaping the architectures of planetary systems. The migration speed and direction of planets depend both on the planet-star mass ratio and the disk's physical properties. 
For low-mass planets (or small planet-star mass ratios), migration is typically described by the viscous Type I regime \citep{Ward1997, Tanaka2002,McNally2019}. This process has been extensively studied, particularly in the context of non-isothermal disks \citep{Paardekooper2010, Paardekooer2011,Ziampras2024a,Ziampras2024b,Ziampras2025}.\\

When a planet becomes massive enough to carve a gap in the disk, migration transitions to the so-called Type II regime \citep[e.g.][]{Lin1979}. In the classical type II migration model, the torques are strong enough to change the density structure of the disk, and the planet opens a deep annular gap around its orbit. However,  hydrodynamical simulations have shown that gas can still drift through the gap \citep[e.g.][]{Masset2001}, challenging the traditional view of no gas flow across it \citep[e.g.][]{Ward1997}. Moreover, this gap-crossing flow allows massive planets to migrate independently of disk accretion \citep[e.g.][]{Duffell2014,DKley2015}. Furthermore, it has been shown that the gap-opening process is influenced not only by the planet’s mass, but also by the disk’s viscosity and scale height \citep[see][]{Crida2006, Duffell2013, Kanagawa2015}. For instance, in a low-viscosity disk (e.g. $\alpha=10^{-4}$, with $\alpha$ a dimensionless measure of the turbulent viscosity in protoplanetary disks), even a Neptune-mass planet can create a gap.
 
 An improved analytical framework to describe the gas structure around gap-opening planets was proposed by \cite{Kanagawa2017}. They showed that a planet interacts mainly with the gas at the bottom of the gap, which is consistent with the results obtained by hydrodynamic simulations. Building on this approach, \cite{Kanagawa2018} created a model for type II migration that is basically type I but reduced due to the lower density inside the gap. They studied the migration of planets with planet-star mass ratios (denoted by $q$) less than $10^{-3}$ and gas viscosities described by the parameter $\alpha=10^{-2}$ and $\alpha=10^{-3}$. On the other hand, \cite{Robert2018} studied the migration of an accreting planet until it reaches the mass of Jupiter, considering lower viscosities values $3 \times 10^{-4} < \alpha <  3 \times 10^{-3}$. They concluded that what drives type II migration is the imbalance between the torques
felt by the planet from the inner and outer disks, as pointed out by \cite{Durman2015}. Moreover, they found that gap-crossing flows are actually negligible at low viscosity, and that cutting this small gas flow with planetary accretion hardly impacts the migration speed.\\

Additional investigation has been conducted for planets that are more massive than the local disk's mass, showing that their migration not only slows down \citep{Syer1995,Ivanov1999}, but can also reverse direction \citep{Crida2007,Hallam2018}. In a recent study, \cite{Dempsey2021} investigated the migration of planets in fixed orbits within the mass range $10^{-3} < q < 2 \times 10^{-2}$, considering viscosity parameters between $10^{-3} < \alpha < 10^{-1}$. Their results revealed cases of positive torques, which were linked to non-negligible disk eccentricities. Building on this, \cite{Scardoni2022} extended the analysis by allowing the planets to change their orbits, and examined the resulting torques. They identified a correlation between migration speed and the parameter $K$, as well as a threshold distinguishing inward from outward migration. Additionally, they emphasized the role of planetary orbital eccentricity in determining the migration direction. Although these studies are based on viscous disk evolution, recent proposals involving wind-driven disk evolution suggest that outward migration may still be possible for massive planets \citep{Kimmig2020}.\\

Most previous studies have assumed high gas disk viscosity values ($\alpha>10^{-3}$). However, recent ALMA observations suggest that turbulence within protoplanetary disks is weak, corresponding to $\alpha<10^{-4}$ \cite[e.g.][]{F2018,F2020,Villaneva2020,Villaneva2022,Pizzati2023}. Additionally, low viscosities may provide a better explanation for certain observed features in protoplanetary disks (see \cite{Rosotti2023}). For instance, in low-viscosity disks, a single planet can create multiple gaps \citep{Bae2017, Dong2017, Dong2018}, even while migrating \citep[e.g.][]{Weber2019}. Moreover, high viscosity values lead to the formation of large protoplanetary disks extending hundreds of au, which have not been observed \citep{Toci2021}. Additionally, compact disks only a few au in size appear to be the most common around low-mass stars ($M_\star<1$~M$_\odot$) \citep{Osmar2025}. Therefore, understanding planet formation around low-mass stars requires a planet migration recipe valid for high-mass planets in low-viscosity disks.\\

Although giant planets are significantly less common around M dwarfs than around Sun-like stars, consistent with expectations from the core accretion model \citep[e.g.][]{Lau2004}, the number of confirmed giant planets orbiting M dwarfs has been steadily increasing. Their occurrence rate is $<1 \%$ \citep[e.g.][]{Bon2013,Sabotta2021}, and so far it has been challenging to reproduce with current planet formation models \citep[e.g.][]{Burn2021}. In particular, their formation becomes difficult when the available dust reservoir is limited to a few Earth masses \citep{Sanchez2024}. Recently, \citet{Stef2023} showed that a Neptune-like planet with an orbital period of just a few days can be formed around a 0.1 M$_\odot$ star, given an initial dust reservoir of 50 M$_\oplus$. Similarly, \citet{Pan2024} demonstrated that giant planets with masses up to that of Saturn may form within 100-day orbits in high-viscosity disks ($\alpha > 10^{-3}$), provided the dust reservoir exceeds 50 M$_\oplus$. Nonetheless, further studies are needed to fully understand their formation as well as migration across different regions of the disk, and to reproduce their occurrence rate. \\

Planet formation models around low-mass stars are hampered by the lack of reliable migration prescriptions for high planet-star mass ratios in low-viscosity disks. Therefore, in this work we explore planetary masses and low-viscosity disks, which are relevant around low-mass stars. We performed hydrodynamical simulations of high planet-star mass ratios ($10^{-3} \leq q \leq  2 \times 10^{-2}$) in low-viscosity disks ($\alpha=10^{-4}$) and explored a wide range of disk physical parameters.
We note that while such planet-star mass ratios correspond to Jupiter-like up to super-Jupiter-like planets around Sun-like stars, they represent planets with masses between 20 M$_\oplus$ and 2 Jupiter masses around low-mass stars of 0.1 M$_\odot$.\\ 

This work is structured as follows. In Section \ref{sec:Method}, we describe the setup of the hydrodynamical simulations and the proposed scenarios of study. In Section \ref{Sec:results}, we present the simulation outcomes, including estimates of total torques and the evolution of planetary orbits. We also provide prescriptions for migration of massive planets in low-viscosity disks and calculate the migration timescales to compare them with exoplanet occurrence rates. In Section \ref{sec:discussion}, we discuss limitations of the model and potential future improvements. Finally, in Section \ref{sec:conclusions}, we summarize our work and highlight key findings.\\


\section{Method}
\label{sec:Method}
To calculate the planet-disk interactions of a single planet embedded in a gas-disk, we used
\textsc{FARGO3D}\footnote{https://github.com/FARGO3D/fargo3d} \citep{Pablo2016}. We set the code to solve the equations of hydrodynamics in two dimensions on an Eulerian polar mesh. As its predecessor \citep[\textsc{FARGO},][]{Masset2000}, the code incorporates fast orbital advection algorithms, allowing for the study of planet–disk interactions over long periods of time. It has been widely used in this field and has demonstrated strong performance across the full range of planetary masses. In particular, it effectively handles the interactions of giant planets which carve gaps in the disk and generate strong shocks in their surrounding \citep[e.g.][]{Valborro2006}.

\subsection{Setup}

We follow the setup from \cite{Kanagawa2018} (hereafter K18). Therefore, the radius unit is determined by the initial location of the planet $r_0$, and the mass unit is determined by the stellar mass $M_\star$. Thus, the surface density unit is $M_\star/r_0^{2}$. We modeled the initial disk surface density $\Sigma(r)$ with a power law\\
\begin{equation}
    \Sigma(r)=\Sigma_0\left(\frac{r}{r_0}\right)^{-s},
    \label{eq:density}
\end{equation}
where $\Sigma_0$ is the gas-disk density at $r=r_0$, with $r$ the radial distance to the star. Similarly, we assumed a power law for the disk aspect ratio\\
\begin{equation}
   h=h_0\left(\frac{r}{r_0}\right)^{f},
   \label{eq:h}
\end{equation}
where $h_0$ is the disk aspect ratio at $r=r_0$. We assumed a steady viscous accretion disk where the factor $\Sigma \nu$ is constant, with the disk viscosity $\nu=\alpha c_s H$, $c_s=\Omega_k H$ the sound speed, $\Omega_k$ the Keplerian frequency, and $H=hr$ the scale height of the disk \citep{Shakura1973}. Following the approach of K18, the density index $s$ is connected to the flaring index $f$ by the relation $s = 2f + 0.5$.\\
We set $\alpha=10^{-4}$ and did not explore lower values, as for low-viscosity disks with $\alpha<10^{-4}$ vortices from the Rossby Wave Instability may alter the planet’s orbit, causing eccentricity growth and erratic inward migration \cite[e.g.][]{McNally2019}. This erratic migration would be hard to capture in a simple torque prescription.\\
We divided the grid into 128 cells in the radial direction (equally spaced in logarithmic
space) and into 512 cells in the azimuthal direction (equally spaced). Thus, one grid cell corresponds to approximately 0.4$H$ for $H/r=0.05$. We note that our resolution is lower than that used in \cite{Kanagawa2018}, as adopting a lower viscosity significantly increases the computational cost at higher resolutions. We applied damping in the wave-killing zones of the disk to avoid artificial wave reflection following \cite{Valborro2006}. The wave-killing zones have a width approximately equal to $10\%$ of the radius of the mesh edges, defined as $r_{in}=0.4r_0$ and $r_{out}=3r_0$, respectively, for the inner and outer edge of the disk. Consequently, the quantity $\Delta r/r=0.015$ and $\Delta \theta=0.012$. Therefore, the number of cells spanning the horseshoe region ranges from 10 to 26 in the radial direction, for mass ratios $10^{-3}<q<2 \times 10^{-2}$.\\
In order to define the gravitational potential of the planet, we used the softening parameter $\epsilon=0.6H_0$. We included the indirect term of the planet, and we modified the code to exclude $60\%$ of the Hill radius of the planet when calculating the force exerted by the disk onto the planet, accounting for the presence of a circumplanetary disk, following \cite{Crida2009}. For simplicity, we neglected the accretion of disk gas onto the planet (see Section \ref{sec:discussion} for discussion on this choice). The planet is allowed to migrate through the disk.

The main difference with K18 is that we consider higher mass ratios, $10^{-3} < q < 2 \times 10^{-2}$, which represent planetary masses between 1 and 20 Jupiter-mass, for a Sun-like star, and between 20 M$_\oplus$ and 2 Jupiter-mass for a star of 0.1 M$_\odot$.

\subsection{Cases of study}

    \begin{table}
    \centering
      \caption{Set of parameters that changed in the simulations.}
     \begin{tabular}{c|c|c|c|c|c}
      q & $h_0$  &  $s$  & $f$ & $\Sigma_0$ & $\alpha$ \\
     \hline
      $[1 \times 10^{-5},   2 \times 10^{-2}]$ & 0.05  &  0.5  & 0 & $10^{-4}$  & $10^{-4}$\\
       $[5 \times 10^{-4},   1 \times 10^{-2}]$ &0.03 &  0.5  &  0 &  $10^{-4}$  & $10^{-4}$\\
       $[3 \times 10^{-3},   2 \times 10^{-2}]$  &0.07  & 0.5  &  0 & $10^{-4}$  & $10^{-4}$\\
       $[3 \times 10^{-3},   2 \times 10^{-2}]$  &0.05  &  1  &  0.25 & $10^{-4}$  & $10^{-4}$\\
       $[3 \times 10^{-3},   2 \times 10^{-2}]$ & 0.05  & 0.5  &  0 &  $10^{-3}$  & $10^{-4}$\\
        $[3 \times 10^{-3},   2 \times 10^{-2}]$ & 0.05  &  0.5  &  0 & $10^{-5}$  & $10^{-4}$\\
       $[3 \times 10^{-3},   2 \times 10^{-2}]$ & 0.05 &  0.5  &  0 & $10^{-6}$  & $10^{-4}$\\
       $[3 \times 10^{-3},   2 \times 10^{-2}]$ & 0.05  & 0.5  &  0 & $10^{-4}$  & $10^{-2}$\\
       $[3 \times 10^{-3},   2 \times 10^{-2}]$ & 0.05  &  0.5  &  0 & $10^{-4}$  & $10^{-3}$\\
        $[5 \times 10^{-4},   1 \times 10^{-2}]$ & 0.03  & 0.5  &  0 & $10^{-3}$  & $10^{-4}$\\
       $[3 \times 10^{-3},   2 \times 10^{-2}]$ & 0.07  &  1  &  0.25 & $10^{-5}$  & $10^{-4}$\\
       \end{tabular}
    \label{tab:cases-of-study2}
\end{table}

\begin{figure*}
    \centering
    \includegraphics[width=0.9\textwidth]{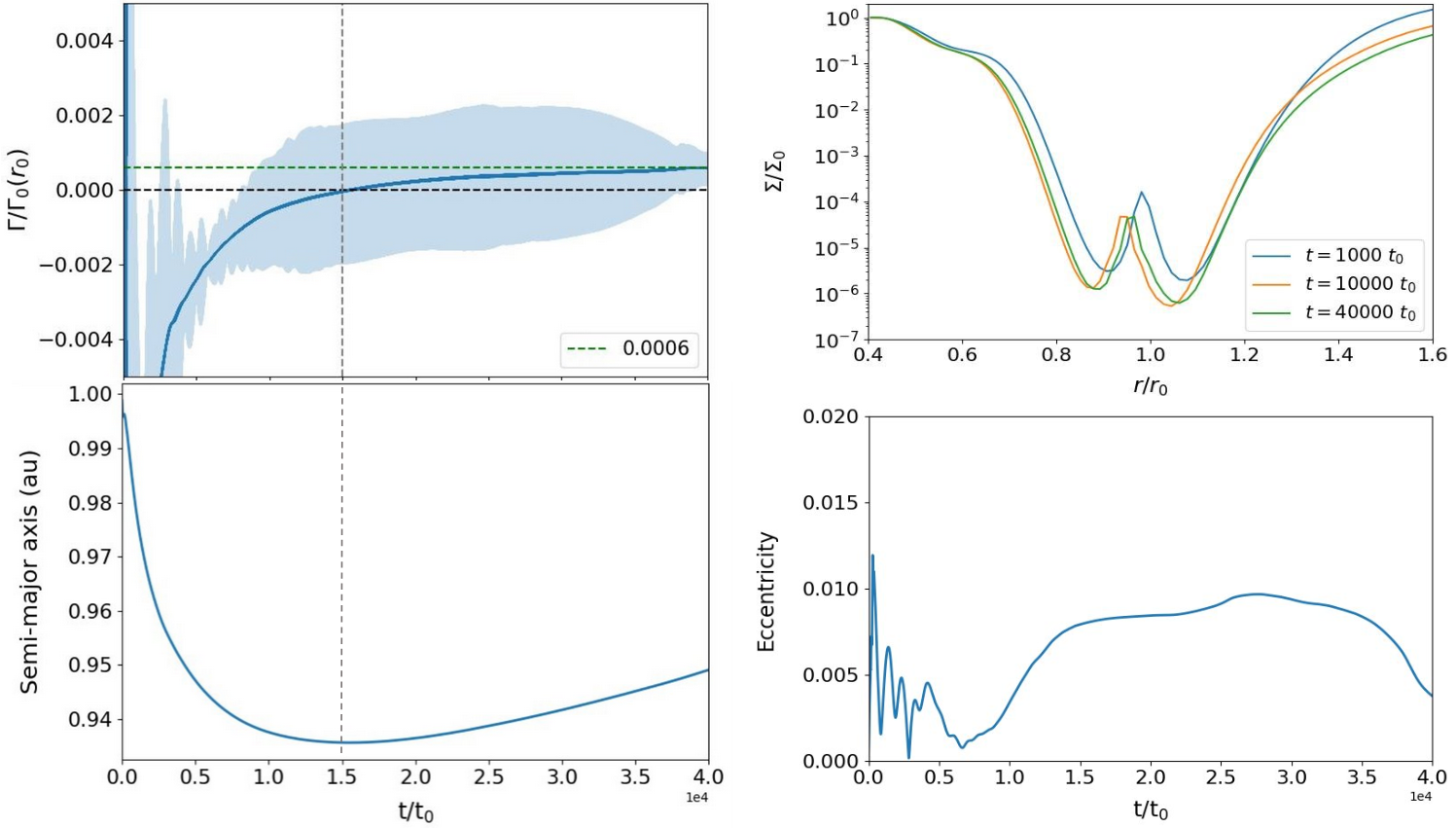}
         
    \caption{Top panels: Evolution of the normalized torque, along with the smoothing function, over the integration (left) for a planet with $q=0.003$ in a low-viscosity disk ($\alpha=10^{-4}$) with $\Sigma_0=10^{-4}$, and $h=0.05$. Distribution of the azimuthally averaged gas-disk density around the planet's initial location at three different integration times (right). Bottom panels: Evolution of the semi-major axis (left) and eccentricity (right) of the planet during the integration. The vertical lines mark the times when the torque changes sign, indicating a reversal in the direction of the planet's migration. Time is expressed in units of $t_0$, the orbital period of the planet at $r=r_0$.}
    \label{fig:standard-torques}
\end{figure*}

 The parameter $K$ that describes the gap depth is associated with the planet-star mass ratio $q$, aspect ratio $h$ and viscosity parameter $\alpha$ by:
\begin{equation}
    K=\left(\frac{M_p}{M_\star}\right)^{2} h^{-5} \alpha^{-1}.
\end{equation}
First, we aim to explore the same range in K-parameter studied by K18, but for migrating planets that do not follow type I migration, associated to $10^{0}< K< 10^{4}$ and embedded in a low-viscosity disk ($\alpha=10^{-4}$). Following the fiducial case in their study, we set $\Sigma_0=10^{-4}$, $h_0=0.05$ and $s=0.5$. The cases studied to recover the same value of $K$ as K18 with our lower value of $\alpha$, are associated to the following planet-star mass ratios $q$: $1 \times 10^{-5}$, $5 \times 10^{-5}$, $1 \times 10^{-4}$, $5 \times 10^{-4}$, and  $1 \times 10^{-3}$.\\

Motivated by the fact that, for a very low-mass star of 0.1 M$_\sun$, the highest value of K explored by K18 (considering $\alpha=10^{-4}$ and $h_0=0.05$) corresponds to a planet with a mass of only 20 M$_\oplus$, we extend the study to higher K values. This expansion allows us to explore the dynamics of more massive planets orbiting very low-mass stars. Thus, we expand our study to $K=2 \times 10^{7}$ which corresponds to a planet with 2 Jupiter-mass around a low-mass star of 0.1 M$_\odot$.

In this new K domain, we explore six different values of planet-star mass ratios $5 \times 10^{-4}$, $1 \times 10^{-3}$, $3 \times 10^{-3}$, $5 \times 10^{-3}$, $1 \times 10^{-2}$, and $2 \times 10^{-2}$. Additionally, we vary several disks parameters, including $0.03<h_0<0.07$, $0.5<s<1$, $10^{-6}<\Sigma_0<10^{-3}$, and $10^{-4}<\alpha<10^{-2}$. The cases of study are listed in Table \ref{tab:cases-of-study2}. We note that the planet is inserted at $t=0$ with its final mass, without employing a growth function to gradually increase the planetary mass over time \citep[e.g.,][]{Griveaud2023}. As a result, the disk undergoes an impulsive perturbation at the beginning of the simulation rather than a gradual adjustment to the planet's presence. However, these initial transients dissipate relatively quickly and do not affect the long-term torque evolution or migration trends.\\
In total, 64 simulations were performed during 40,000$t
_0$, with $t_0$ the orbital period at $r=r_0$. We note that the simulations need to be run for an extended period because the viscous timescale of the disk, which determines how quickly the disk reacts to the planet’s presence, is typically much longer in low-viscosity disks. As a result, a longer simulation duration is necessary for the disk to adjust and for a stable interaction between the planet and the disk to be reached. We also point out that in the simulations, the planet migrates only a short distance from its initial location (less than $\sim5\%$). As it remains close to its starting point, even in the case of $f\neq0$, the value of $h$ stays approximately constant, and so does the corresponding value of $K$.

\section{Results}
\label{Sec:results}
We detail the procedure used to estimate the torques exerted by the gas disk on the planet, based on the outcomes of our hydrodynamical simulations. Additionally, we present the resulting torques for the different cases of study proposed in the previous section. Furthermore, we introduce new prescriptions to account for the expanded parameter space explored in this work. Finally, we estimate migration timescales and compare them with occurrence rates of giant planets.

\begin{figure*}
    \centering
    \includegraphics[width=0.47\textwidth]{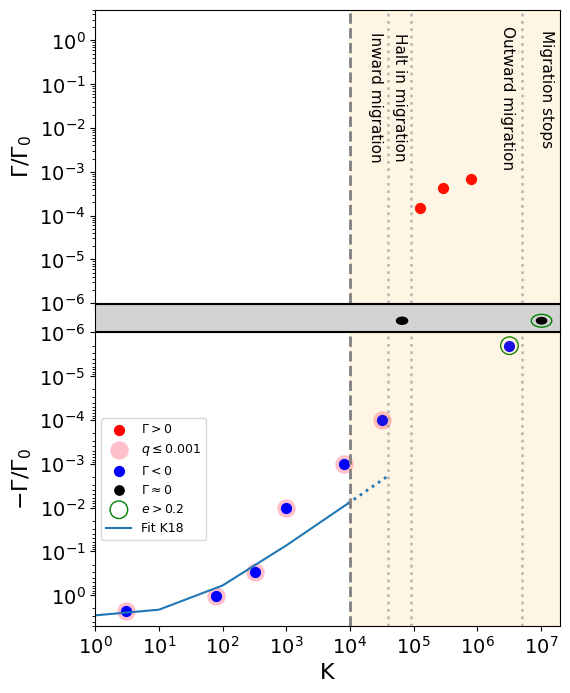}
    \includegraphics[width=0.48\textwidth]{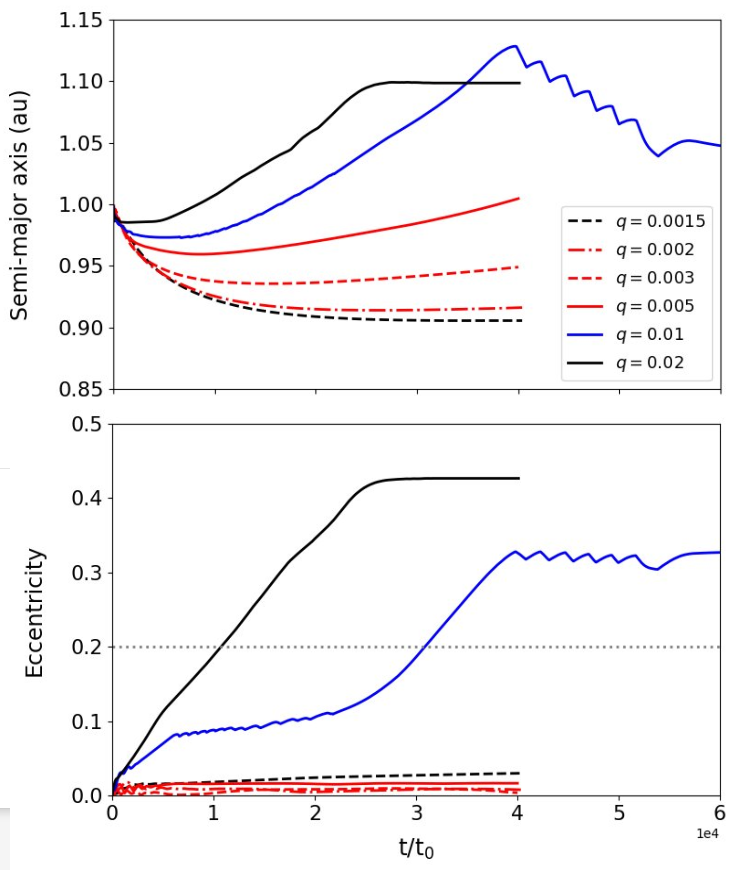}
    \caption{Left: Normalized positives (red dots), negatives (blue and pink for $q \leq 1 \times 10^{-3}$), and close to zero (black) torques as a function of a wide range of the $K$-parameter, including the previously studied domain by K18 (white background) and the new domain explored in this work (cream background), assuming $\alpha=10^{-4}$, $h_0=0.05$, $s=0.5$, and $\Sigma_0=10^{-4}$. The fitting proposed by K18 in the $K$-range of their study (solid blue line) together with the potential extension in the new regime (dotted blue line) is given in Eq. \eqref{eq:K2018-fit}. Torques exerted onto planets in orbits with $e>0.2$ are highlighted (green circles). Four different migration regimes are marked indicating inward migration, a halt in migration, outward migration, and migration stalls (dotted gray lines). Right: Evolution of the semi-major axis (top) and the  eccentricity of the six planets that experience either outward migration or a halt in migration: $q=1.5 \times 10^{-3}$ (dotted black line), $q=2 \times 10^{-3}$ (dashed-dotted red line), $q=3 \times 10^{-3}$ (dashed red line), $q=5 \times 10^{-3}$ (solid red line), $q=1 \times 10^{-2}$ (solid blue line) and $q=2 \times 10^{-2}$ (solid black line). The value of eccentricity $e=0.2$ that gives the change in the torques sign is overplotted (gray dotted line). The simulation for the planet with $q=0.01$ was extended to allow eccentricity to settle.}
    \label{fig:total-torques}
\end{figure*}

\subsection{Quantifying torques and planetary dynamics}
\label{sec:torque-descrip}

In order to determine the torque that the gas-disk exerted onto the planet, we integrated our simulations for around 40,000 orbits. Even for these long integration times, the low-viscosity ($\alpha=10^{-4}$) nature of the disk means that true steady states are hard to reach. For quantifying the torques, we employed a Savitzky-Golay filtering that applies a polynomial smoothing technique to the data, which smooths data while preserving features such as peak height, width, and slope better than a running average \citep{Savitzky1964}. \\
As an example, in Figure \ref{fig:standard-torques} we show, for a mass ratio $q=3 \times 10^{-3}$, the evolution of the torque measured in the simulation $\Gamma$ normalized to the scaling torque $\Gamma_0$, which is defined by:
\begin{equation}
    \Gamma_0(r)=\left(\frac{M_p}{M_\star}\right)^{2}h^{-2} \Sigma(r) r^{4} \Omega_k^{2}.
    \label{eq:normalized_torque}
\end{equation}
We also include the corresponding smoothed fitting function obtained from the torque data to show the trend in the torque evolution. We can see that the torque changes sign after approximately 15,000 orbits, becoming positive and approaching the value $\Gamma/\Gamma_0 \approx 0.0006$. Although the torque did not reach a steady state within the integration period, it appears unlikely that the torque will undergo any qualitative changes.\\

Additionally, we analyze the distributions of the azimuthally averaged gas surface density, normalized by the initial unperturbed local density of the planet, alongside the evolution of the semi-major axis and eccentricity of the planet throughout the integration. As an example, Figure \ref{fig:standard-torques} also illustrates these three quantities for the case of $q=3 \times 10^{-3}$. It shows that the position of the gap carved by the planet in the density profile shifts according to the torque’s sign, moving either inward toward the star or outward. Furthermore, since the eccentricity remains small ($e<0.015$), the negative torque corresponds to inward migration, as indicated by the decreasing semi-major axis over time. Once the torque becomes positive, the planet reverses direction and begins to migrate outward.\\

\begin{figure*}
    \centering
  \includegraphics[width=0.275\textwidth]{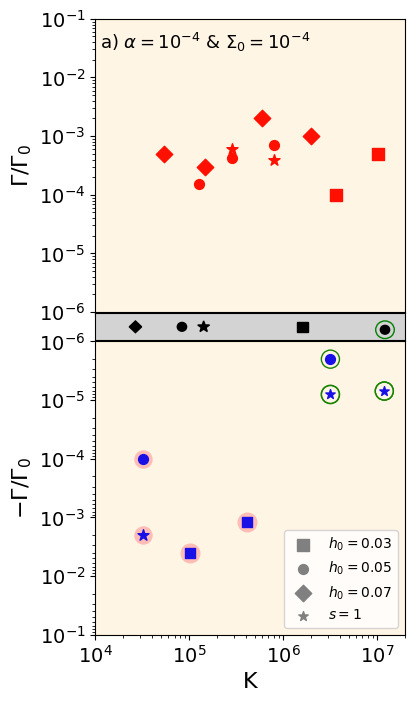}      \includegraphics[width=0.23\textwidth]{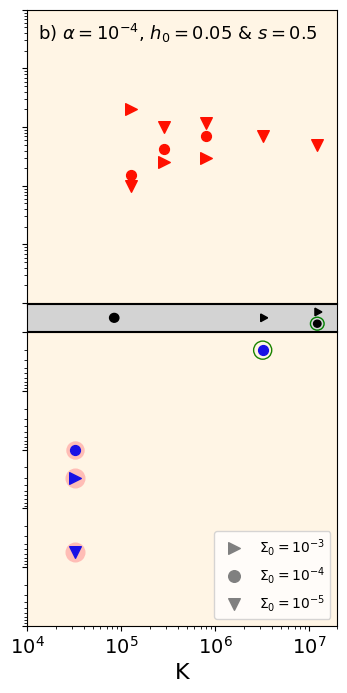}
    \includegraphics[width=0.23\textwidth]{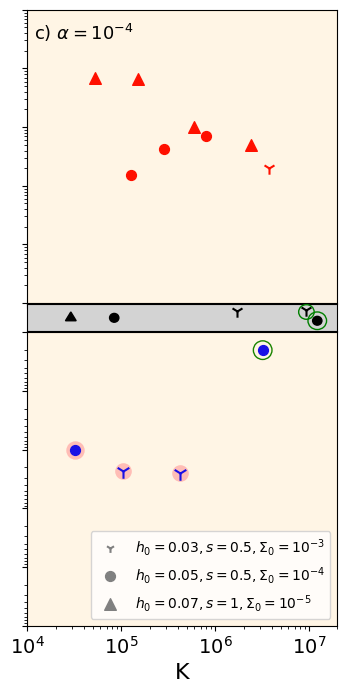}
    \includegraphics[width=0.23\textwidth]{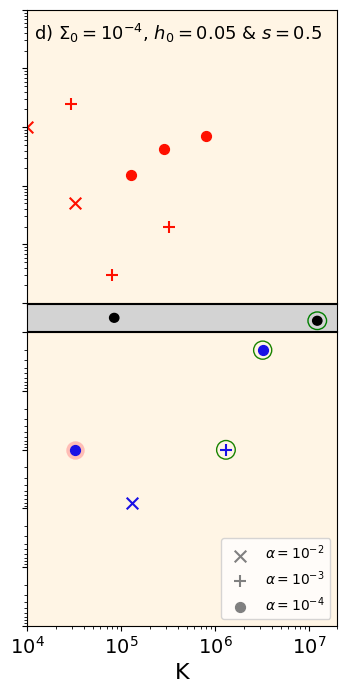}\\
    
    \caption{Normalized torques vs the K-parameter range proposed in this study for the mass ratio $5 \times 10^{-4}  < q < 2 \times 10^{-2}$ a) with fixed $\alpha=10^{-4}$ and $\Sigma_0=10^{-4}$, varying $h_0$ or $s$; b) with fixed $\alpha=10^{-4}$, $h_0=0.05$, and $s=0.5$, varying $\Sigma_0$; c) by changing $h_0$, $s$ and $\Sigma_0$ to simulate realistic conditions along the disk; d) with fixed $\Sigma_0=10^{-4}$, $h_0=0.05$, and $s=0.5$, varying $\alpha$. Same color palette as in Figure \ref{fig:standard-torques}.}
    \label{fig:torques-casesofstudy}
\end{figure*}

\begin{figure*}
    \centering
    \includegraphics[width=0.95\linewidth]{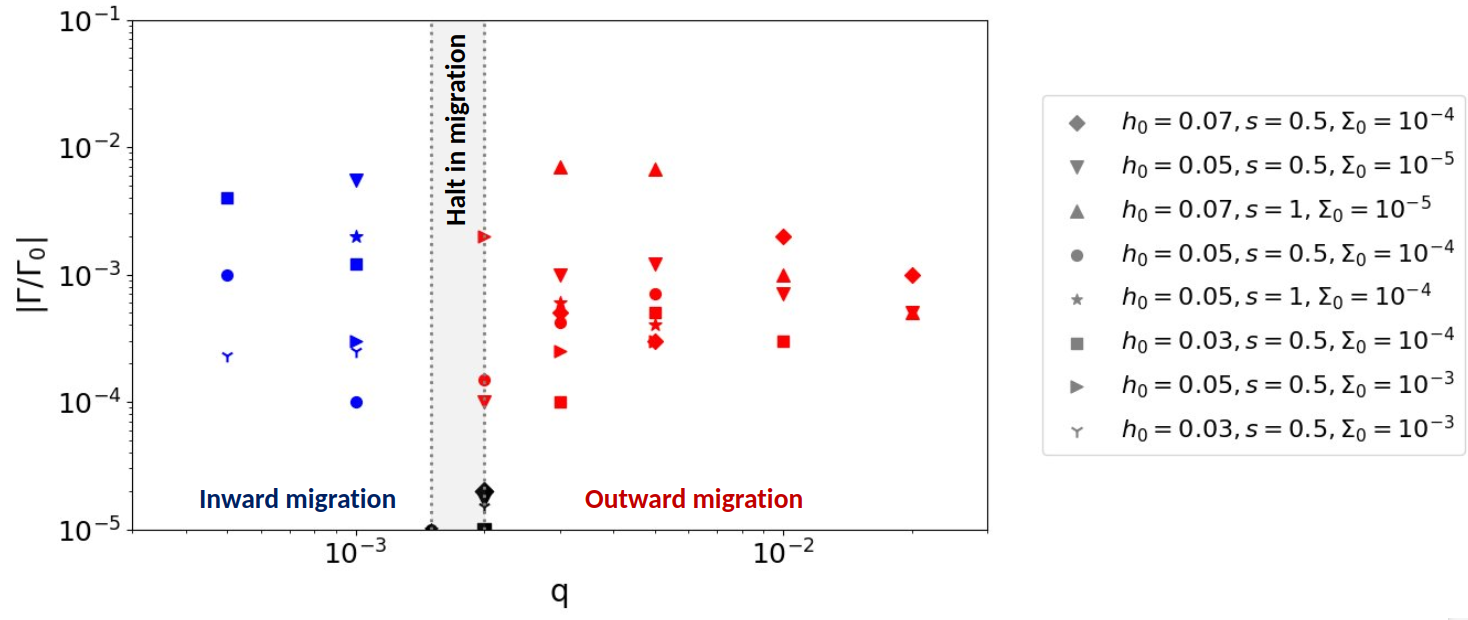}
    \caption{Normalized torques for the different mass ratios associated with planets in low-viscosity disks ($\alpha=10^{-4}$) and with orbital eccentricities $e<0.2$. The transition between inward (blue symbols) and outward (red symbols) migration (shaded area) occurs for $q$ between 0.0015 and 0.002 (black dots), depending on the disk's physical parameters. Symbols and colors are the same as in Figure~\ref{fig:torques-casesofstudy}. For cases with near-zero torque, values around $1 \times 10^{-5}$ were used for better visibility in the figure.}
    \label{fig:torques-q}
\end{figure*}

\begin{figure*}
    \centering
    \includegraphics[width=0.98\textwidth]{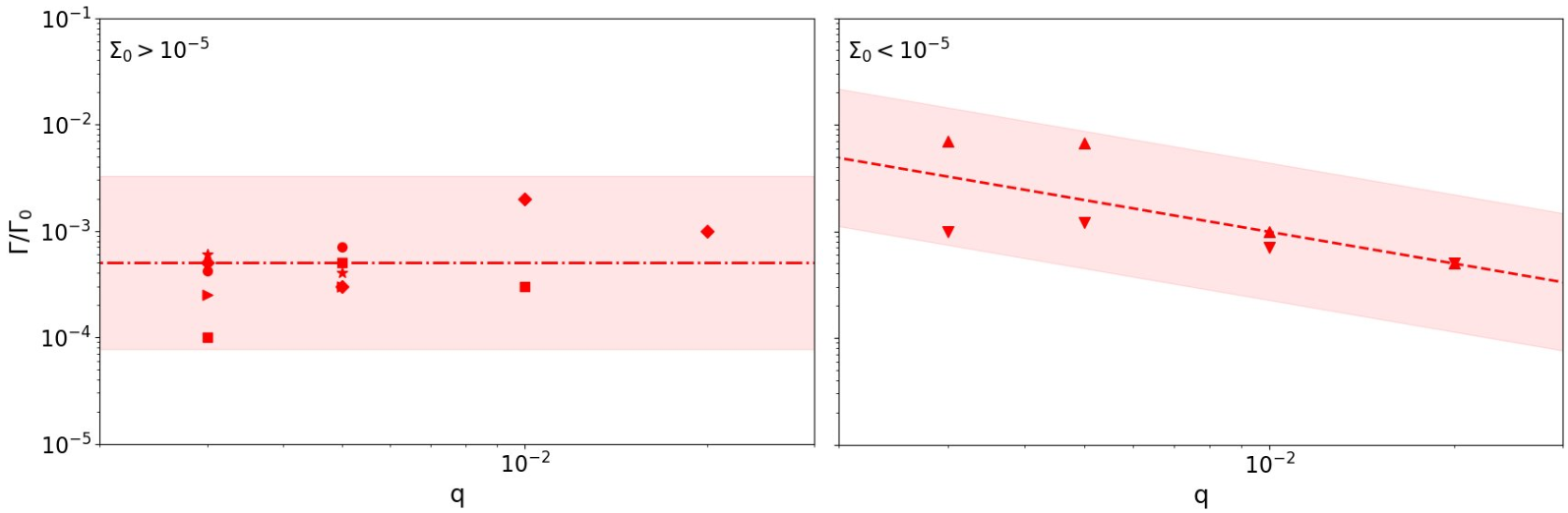}    
    \caption{Absolute values of the normalized torques for planets undergoing outward migration in low-viscosity disks ($\alpha=10^{-4}$) with $2\times10^{-3} < q < 2\times10^{-2}$ and $e<0.2$ (see Table \ref{tab:cases-of-study2}). Left: results from the simulations assuming a local density $\Sigma_0>10^{-5}$. Right: results from the simulations assuming $\Sigma_0<10^{-5}$. In each panel, the fitting function proposed in this work (dashed red lines-see Eq. \ref{eq:newfit}) along with a region within $\pm 2$ the standard deviation of the logarithmic residuals (red shadow area). Symbols and colors are as in Figure~\ref{fig:torques-casesofstudy}.}
    \label{fig:Newprescriptions}
\end{figure*}

\begin{figure}
    \centering
    \includegraphics[width=1\linewidth]{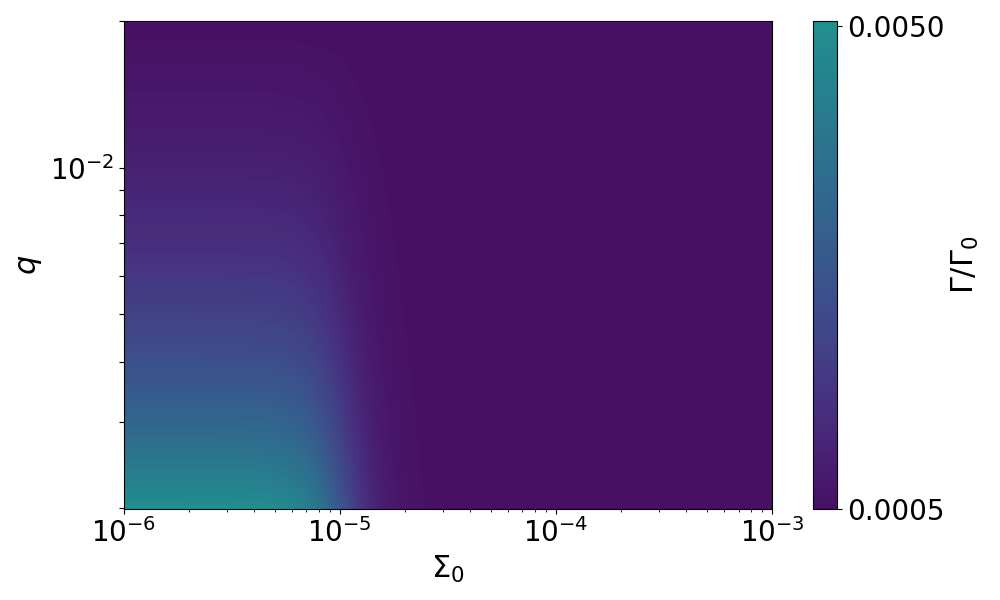}
    \caption{Normalized torque map for the different values of mass ratios ($q$) and local densities ($\Sigma_0$) explored in this work fow planets in low-viscosity disk ($\alpha=10^{-4}$) with eccentricity values $e<0.2$.}
    \label{fig:density-q-map}
\end{figure}

 We present the normalized torques from simulations with low viscosity ($\alpha=10^{-4}$), and standard values $\Sigma_0=10^{-4}$, $h_0=0.05$ and $s=0.5$ in Figure \ref{fig:total-torques}. The torques are categorized as negative, positive, or those remaining close to zero. For the same K interval explored by K18 ($10^{0} < K < 10^{4}$), our results show good agreement, with the torque values aligning well with their fitting function:
\begin{equation}
    \frac{\Gamma}{\Gamma_0}=-\frac{c}{1+0.04K},
    \label{eq:K2018-fit}
\end{equation}
with the fitting constant $c=3$. In principle, it seems like the fitting function could be applied up to $K=4 \times 10^{4}$. Beyond this value, there is a narrow range where migration may halt (represented in this case by $q=0.0015$), and for $K$ between $10^{5}$ and $10^{6}$, we observe positive torques consistent with outward migration. For the highest K values (up to $K=2 \times 10^{7}$), associated with $q=0.01$ and $q=0.02$, we find either slight outward migration or negligible migration. When the torques are positive and the planet is migrating outwards, the eccentricity remains close to zero. However, for the highest mass ratios considered, we observe that the planet's eccentricity can reach values in the range $0.2 < e < 0.45$. In particular, for the case of $q = 0.01$, we observe a negative torque associated with outward migration and a growing eccentricity up to around 40,000 orbits. To better characterize the migration path, we extended the simulation by an additional 20,000 orbits. We find that the planet’s eccentricity peaks at approximately $e \sim 0.3$, followed by small oscillations. These oscillations are also reflected in the evolution of the semi-major axis, which gradually decreases until both parameters eventually stabilize over time.
These results confirm that planets with high mass ratios $q$ can experience eccentricity growth up to about 0.45, while remaining gravitationally bound to their host star. The evolution of the semi-major axis and eccentricity for the cases of outward or negligible migration is illustrated in Figure~\ref{fig:total-torques}. \\

\subsection{Measured torques and their dependence with $K$}

To gain deeper insight into migration trends at high $K$ values for mass ratios $5\times10^{-4} \le q \le 2\times10^{-2}$, we analyzed torque values under a range of disk conditions, as summarized in Table \ref{tab:cases-of-study2}. Figure \ref{fig:torques-casesofstudy} presents the normalized torques obtained for different combinations of $h_0$, $s$, $\Sigma_0$, and $\alpha$.

As shown in Figure \ref{fig:torques-casesofstudy}, for local surface densities $\Sigma_o>10^{-5}$, the normalized positive torque values are scattered within the same order of magnitude across the explored range of $h_0$. Additionally, we found qualitatively similar results when varying the density slope from $s=0.5$ to $s=1$ ($h_0=0.05$). In the cases explored in this work, the planets exhibit slow migration timescales. However, we acknowledge that, as shown by \cite{Paardekooper2010,Paardekooer2011}, the total torque—particularly the corotation component—can depend strongly on the local density slope, independently of the planet’s orbital evolution. The weak dependence observed in our simulations likely reflects the fact that the corotation torque is strongly diminished in the gap-opening regime, causing the total torque to become less sensitive to the local surface density slope.

In contrast, for a lower surface density $\Sigma_0=10^{-5}$, the normalized torque strength increases. However, examining the non-normalized torques reveals that their absolute values decrease with decreasing local disk density, consistent with slower migration rates as expected from \citet{Baruteau2014}. Additional simulations with $\Sigma_0=10^{-6}$ and $\Sigma_0=10^{-7}$ produced normalized torque values nearly identical to those obtained with $\Sigma_0=10^{-5}$, reinforcing this trend. These cases are not shown in Figure \ref{fig:torques-casesofstudy} to avoid over-plotting, as the resulting values would be indistinguishable from those already displayed.

We note that, for cases with $\Sigma_0 \ge 10^{-4}$ and $h_0 \le 0.05$, planets with mass ratios $q = 0.01$ or $q = 0.02$ tend to reach eccentricities in the range $0.2 < e < 0.45$, while the absolute torque values remain negative or close to zero. In particular, after 40,000 orbits, planets with $q = 0.01$ which exhibit negative torques during the phase of eccentricity growth, are expected to stop migrating once the eccentricity stabilizes  (see Subsection \ref{sec:torque-descrip}). These results highlight the significant influence of planetary eccentricity on migration behavior. Notably, even when torques are negative, outward migration can still occur if the planet follows an eccentric orbit, as it is the total work done by the disk that ultimately determines the migration direction \citep{Cresswell2007,2024arXiv241000374B}. 
Further analysis of the 2D gas density map reveals a pronounced asymmetry in the gap carved by the planet, indicating a growing eccentricity in the disk itself. This disk eccentricity may contribute to the evolution of the planet’s eccentricity. These findings are consistent with previous studies \citep{Papa2001,Goldreich2003,Kley2006}, which show that eccentricity growth is favored in disks with low aspect ratios and low viscosities, particularly for gap-opening planets.\\

On the other hand, when exploring higher values of $\alpha$ while keeping $\Sigma_0=10^{-4}$, $h_0=0.05$, and $s=0.5$, we measured positive torques for all planets with $q<0.01$, consistent with results reported by \citet{Dempsey2021}. However, for the highest mass ratio considered in this study ($q=0.02$), the torques remained negative. When the eccentricity stayed below $e<0.2$, positive torques are related to outward migration, while negative torques corresponded to inward migration. We also observed that, within the explored $K$ range, higher viscosities introduce greater scatter in the absolute torque values, and the direction of migration becomes less stable at the highest mass ratios. In contrast, for low viscosities, outward migration persists up to the largest mass ratios considered, although it may slow or stall rather than reverse direction.

Overall, across all our simulations assuming a low-viscosity disk, we do not find a straightforward relation between the strength of the torque and the $K$-parameter. 
Interestingly, we observe a halt in migration that marks a transition in migration direction, associated with a mass-ratio $q \approx 0.002$, depending on the physical properties of the disk. This is explored in more detail in the following section.

\subsection{Migration reversal at mass-ratio $q \approx 0.002$}

To reinforce the robustness of the migration direction change at $q \approx 0.002$ and its independence from the $K$-parameter, we ran additional simulations with planets with $q=0.002$ embedded in low-viscosity disks ($\alpha=10^{-4}$) characterized by aspect ratios $h_0$ from $0.03$ to $0.2$, covering a wide range of $K$ values from $10^{2}$ to $10^{7}$. As a result, as long as $K>10^{4}$, we consistently found that planetary migration either stalls or becomes slightly outward at $q=0.002$. However, for $K<10^{4}$ we recovered inward migration as in K18. 

We realized that at a mass ratio $q=0.002$, migration behavior is highly sensitive to the disk’s physical properties. For $h_0=0.05$, the torque becomes positive, indicating outward migration, while for $h_0=0.03$ and $h_0=0.07$, the torque is nearly zero at the same mass ratio, assuming a surface density slope $s=0.5$. However, when the surface density slope steepens to $s=1$, in the case of $h_0=0.05$, the simulation shows migration stalls at this mass ratio (not at $q=0.0015$ as in the default scenario with $s=0.5$). This sensitivity indicates that planets with mass-ratios $q$ between 0.0015 and 0.002 reside in a nonlinear migration regime where small variations in disk structure—such as scale height and surface density slope—can significantly alter the torque balance by affecting Lindblad and corotation torques. This highlights the complexity of migration in low-viscosity disks, where small structural changes can have outsized impacts on the migration direction and rate.

This led to the emergence of a robust and consistent pattern:
\begin{itemize}
\item Planets with $q<1.5\times10^{-3}$ for $K>10^{4}$ consistently experience inward migration.\\
\item Planets with $q$ between 0.0015 and 0.002 may experience a halt in migration, depending on the disk’s physical properties.\\
\item Planets with $2\times10^{-3}<q<2\times10^{-2}$ for $K<2\times10^{7}$ undergo outward migration, provided their eccentricities remain below $e<0.2$.
\end{itemize}
This transition in migration direction is illustrated in Figure~\ref{fig:torques-q}, which compiles all normalized torques estimated for planets in low-viscosity disks with $e < 0.2$.\\

Furthermore, for planets undergoing outward migration in low-viscosity disks ($\alpha = 10^{-4}$) with $e < 0.2$, we found that the torque strength correlates with the local gas surface density according to the following relation:
\begin{equation}
\label{eq:newfit}
\Gamma/\Gamma_0 = 
\begin{cases}
 ~~a, & \text{if } ~~\Sigma_0 > 10^{-5} \\
 ~~ \frac{b}{q},  & \text{if }~~ \Sigma_0 \le 10^{-5}
\end{cases}
\end{equation}
with $a=5\times10^{-4}$ and $b=10^{-5}$.  These fits are shown in Figure~\ref{fig:Newprescriptions}, along with the measured normalized positive torques driving outward migration for planets with $2\times10^{-3} < q < 2\times 10^{-2}$. Notably, for each fitting prescription, all measured torques fall within two standard deviations of the logarithmic residuals ($\sigma = 0.44$ dex and $\sigma = 0.32$ dex, respectively), demonstrating how well the fit reflects the torques estimated from the simulations.

To avoid a discontinuity at $\Sigma_0 = 10^{-5}$, we further propose the following function:
\begin{equation}
    \Gamma_{\rm {fit}}=\Gamma/\Gamma_0 = a(1-f_{\rm {fit}}) + (b/q)f_{\rm{fit}},
    \label{eq:smoothf}
\end{equation}
where $f_{\rm{fit}}=1 / (1 + (\Sigma_0/ 10^{-5})^5)$ is a smooth transition function designed to preserve the same torque values as for $\Sigma_0=10^{-5}$ in the low-density regime ($\Sigma_0 < 10^{-5}$). Figure~\ref{fig:density-q-map} shows the application of this function to estimate normalized torques across the range of mass ratios and local gas densities studied in this work. We note that for the intermediate mass ratios $q\sim0.003-0.005$, lower disk surface densities lead to stronger (more positive) normalized torques due to enhanced gap asymmetry. At higher mass ratios $q > 0.005$, the planet dominates the local dynamics, making the torque less sensitive to the disk mass. This explains the convergence of normalized torques at high $q$ across different disk densities.\\

These results offer a consistent framework to interpret outward migration trends in low-viscosity disks, with potential applications in population synthesis and planet formation models.

\subsection{Estimation of migration timescales across the disk}

\begin{figure*}
    \centering
    \includegraphics[width=0.93\textwidth]{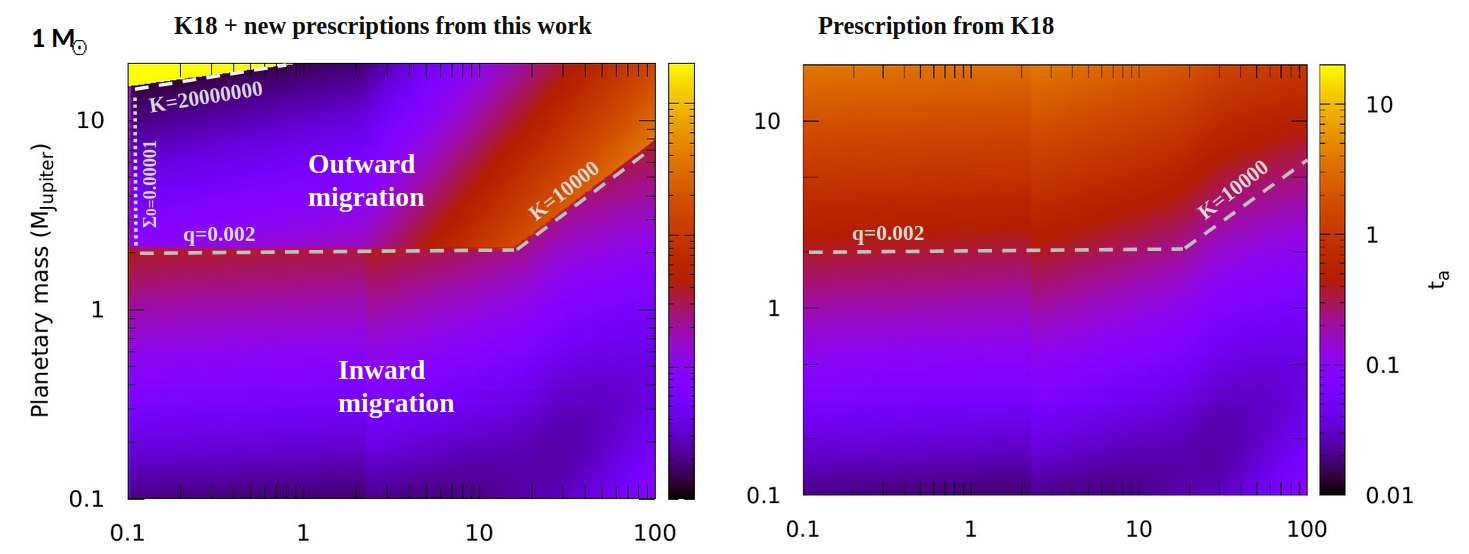}\\
     \includegraphics[width=0.95\textwidth]{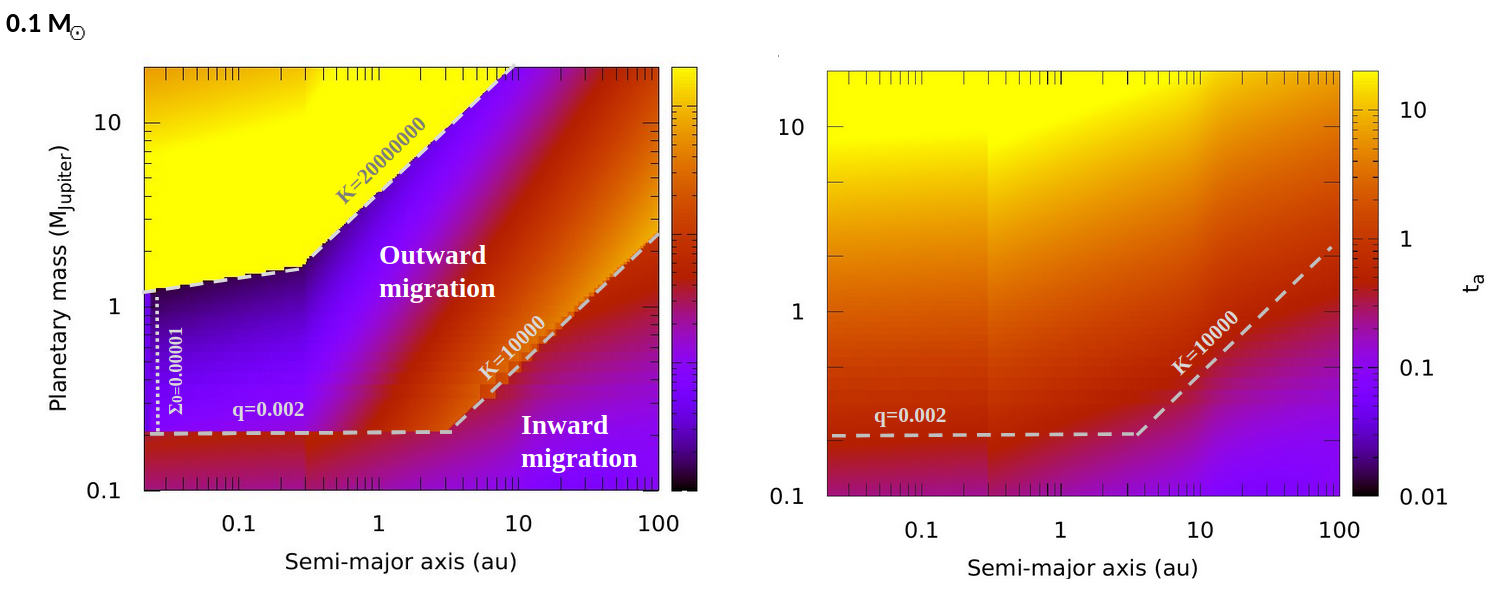}
    \caption{Maps of semi-major axis evolution timescale for different planetary masses and semi-major axis for a 1 M$_\odot$ star (top) and for a 0.1 M$_\odot$ star (bottom), assuming a low-viscosity disk ($\alpha=10^{-4}$) with gas surface density and aspect ratio profiles described in Eqs. \eqref{eq:density} and \eqref{eq:h}, assuming $s=0.5$ ($f=0$) in the inner disk and $s=1$ ($f=0.25$) in the outer disk. The transition was set at $r_{\rm{trans}}=0.3$~au for $M_\star=0.1$~M$_\odot$, and at $r_{\rm{trans}}=2$~au for $M_\star=1$~M$_\odot$. We assumed $r_0=r_{\rm{tran}}$, and characterized both the inner and outer disk by 
$\Sigma_0=4 \times 10^{-4}$~M$_\odot/au^{2}$ and 
$h_0=0.03$, simulating the conditions of a disk at 1–2 Myr with a high gas disk mass $\sim 10\%$~M$_\star$. Limits for inward and outward migration are overplotted (dashed gray lines). Right: timescales calculated applying only the fitting function from K18 (see Eq. \eqref{eq:K2018-fit}). Left: timescales calculated combining the fitting function from K18 for $K<10^{4}$ and $q<0.002$, and the fitting functions propose in this work (see Eq. \ref{eq:newfit}), for $0.002<q<0.02$ and $K<2\times10^{7}$, relating the surface density $\Sigma(r)$ to $\Sigma_0=10^{-5}$~M$_\star/r^{2}$ (dotted gray line)}.
    \label{fig:ta-01-1Msol}
\end{figure*}

We aim to quantify migration for planets with different masses, within the range considered in this study, embedded in a low-viscosity disk ($\alpha = 10^{-4}$). Since our analysis focuses on planets in nearly circular and coplanar orbits, the migration timescale, defined as the rate of change in angular momentum, can be expressed as $\tau_m \approx 2\tau_a$, where $\tau_a$ denotes the timescale associated with the change in semi-major axis:
\begin{equation}
    \tau_a=-\frac{a}{da/dt}=\frac{a^2\Omega_k M_p}{2\Gamma}.
\end{equation} 

Therefore, we computed $\tau_a$ for giant planets throughout the disk, following the normalized torque fitting function proposed by K18 (see Eq. \eqref{eq:K2018-fit}) for planets migrating inward, and the fitting function introduced in this work (see Eq.\eqref{eq:smoothf}) for planets migrating outward. We computed the scaling torque using the gas surface density and aspect ratio profiles described in Eqs.~\eqref{eq:density} and \eqref{eq:h}, assuming $s=0.5$ ($f=0$) in the inner disk to simulate a viscous disk, and $s=1$ ($f=0.25$) in the outer disk to represent an irradiated disk \citep[e.g.,][]{Ida2016}. We assume $r_0=r_{\rm{tran}}$, the transition radius between the inner and outer disk. For a star of $M_\star=0.1$~M$_\odot$, this transition is set at $r_{\rm{tran}}=0.3$~au, while for 
$M_\star=1$~M$_\odot$ it is set at $r_{\rm{tran}}=2$~au. In both cases, we adopted inner and outer disk profiles characterized by 
$\Sigma_0=4 \times 10^{-4}$~M$_\odot/au^{2}$ and 
$h_0=0.03$, simulating the conditions of a disk at 1–2 Myr with a high gas disk mass $\sim 10\%$~M$_\star$. 

Figure \ref{fig:ta-01-1Msol} presents maps of the absolute value of $\tau_a$ across different planetary masses and semi-major axes for stars of 0.1 M$_\odot$ and 1 M$_\odot$. For comparison, these maps include the results of applying the new migration patterns introduced in this work, along with those considering only the inward migration model proposed by K18. We note that the disk density will be smaller than $10^{-5}$~M$_\star/r^{2}$  for large radial distances ($r>100$~au), but also for a small region close to the inner edge of the disk. Near the inner edge of the disk, the surface density at the planet's location typically exceeds $10^{-3}$~M$_\odot/au^2$. For example, for a $M_\star=0.1$~M$_\odot$ and $r=0.025$~au, this corresponds to $\sim 9 \times 10^{-6}$~$M_\star/r^2$ in code units—just below the $10^{-5}$ threshold. We also ascribe to the fact that the upper limit for outward migration in our study is defined by the highest $K$-value examined ($K = 2 \times 10^{7}$). As migration appears to slow down with increasing $K$, we associate the region with $K > 2 \times 10^{7}$ with slower migration rates. However, this is linked to even higher mass ratios, and a further study, which fall outside the scope of this work, is needed to properly estimate the timescales. \\
Regardless of the stellar mass, we observe that the new normalized torques estimated in this work allow giant planets to experience different migration speeds and directions. In particular, for planets with masses $M_{p}>2$ Jupiter mass for a star of 1 M$_\odot$, and for planets with masses $M_p>0.2$ Jupiter mass for a star of 0.1 M$_\odot$. \\

These new findings on outward migration in low-viscosity disks provide a new perspective on how giant planets may form throughout the disk.

\subsection{Linking giant planet occurrence rates and migration timescales}

\begin{figure}
    \centering
    \includegraphics[width=1.\linewidth]{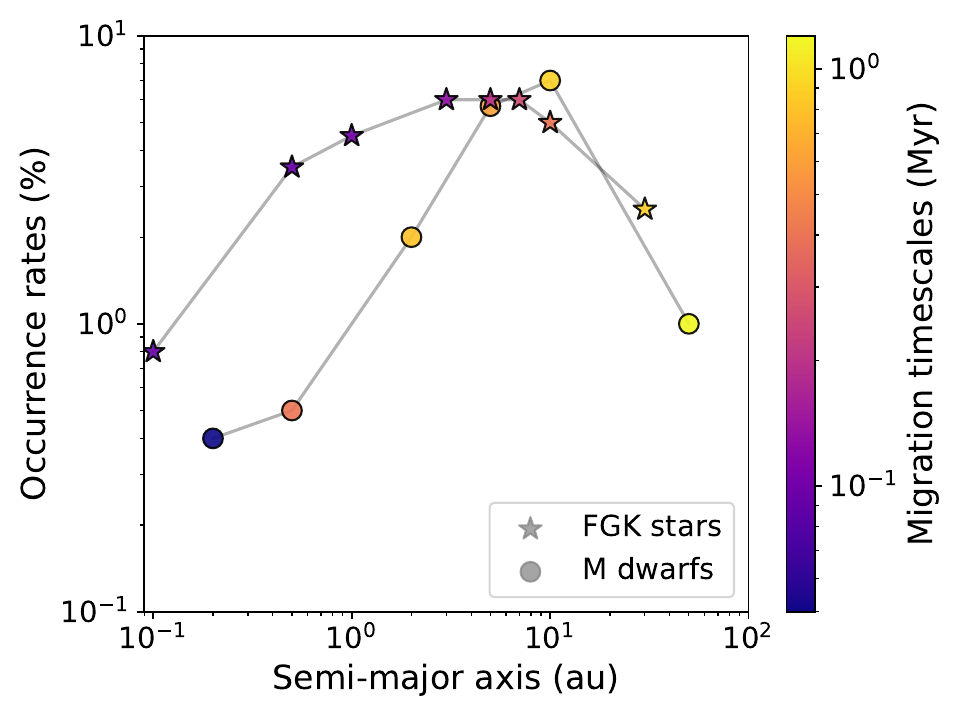}
    \caption{Occurrence rates of giant planets with masses between 0.1 and 20 Jupiter-masses as a function of semi-major axis, correlated with migration timescales at each location, for FGK stars and M dwarfs. Quasi-circular orbits are assumed for the planets.}
    \label{fig:OR-ta}
\end{figure}

Motivated by the varying patterns of migration across the disk, we seek to qualitatively compare the occurrence rates of giant planets with their estimated migration timescales, which are associated to the time it takes for a planet to undergo significant orbital migration due to interactions with the surrounding protoplanetary disk. Specifically, we compared occurrence rates of giant planets with migration timescales for stars of 1 M$_\odot$ and 0.1 M$_\odot$, respectively. For FGK stars, we adopt the occurrence rates of planets with masses between 0.1 and 20 Jupiter masses from \cite{Fernandes2019}, derived from a log-normal distribution ($\mu=3.16$~au, $\sigma=0.75$~dex) that fits Kepler data \citep{Bergsten2022}, as well as radial velocity \citep{Fulton2021} and direct imaging data \citep{Vigan2017}. For M dwarfs, we adopted the occurrence rates of giant planets with masses between 1 and 10 Jupiter masses by integrating the maximum-likelihood log-normal distribution proposed by \citet{Meyer2018}, which compiles results from various observational techniques. To ensure completeness, we also included the estimates from \citet{Clanton2014}, which combine constraints from microlensing and radial velocity for planets between 0.1 and 13 Jupiter masses, spanning orbital periods between 100 and 10,000 days (corresponding to semi-major axes between 0.2 and 5 au). Additionally, occurrence rates for planets with masses between 0.1 and 1 Jupiter mass around M dwarfs at separations $a<1$~au (assuming $M_\star=0.1$~M$_\odot$) have also been reported, and remain low, below 0.16$\%$ \citep{Sabotta2021}. 

In Figure \ref{fig:OR-ta}, we present the occurrence rates as a function of semi-major axis for FGK stars and M dwarfs, correlated with the average migration timescale at each location, within the mass range described above, which is associated with the calculation of occurrence rates. We note that planets located at $a<1$~au exhibit the fastest migration, consistent with the lowest occurrence rates estimated in previous works. Additionally, there is an increase in migration timescales for planets located between 3 and 10 au, which corresponds with higher occurrence rates. We emphasize that, at least for the inner disk region (in this example, $a < 0.3$~au around the M dwarf and $a < 2$~au around the Sun-like star), where the surface density slope is $s = 0.5$, the migration timescale ($\tau_m$) is independent of the planet’s orbital distance from the star. Therefore, any variation in migration timescale across the inner disk reflects genuine physical differences in the torque acting on the planet, rather than being a simple consequence of orbital period scaling.\\
On the other hand, the estimated migration timescales for planets beyond $a>10$~au are still high, which contradicts the observed drop in occurrence rates at these locations. One possible explanation for this discrepancy is that migration timescales are typically calculated under the assumption of quasi-circular orbits, whereas most giant planets beyond 10 au exhibit high eccentricities\footnote{https://exoplanetarchive.ipac.caltech.edu/index.html}. Since eccentric orbits lead to faster migration 
this could help reconcile the lower occurrence rates observed at these distances, considering just the planets with $q<0.02$. Another possible explanation is that giant planet formation becomes inefficient beyond 10 au. As suggested in previous studies, core formation timescales increase significantly at larger distances from the star \citep[e.g.][]{Morbidelli2016}. Furthermore, if planet formation could take place at such location, the low migration timescales could keep the planet fixed at wider orbits, and provide an explanation for planet-inferred gaps in protoplanetary disks \citep[e.g.][]{Lodato2019,Nienke2023,Osmar2025}. Overall, the fact that super-Jupiters  migrate outward could explain some of the directly imaged planets in the 10-100 au range \citep{Vigan2017,Nielsen2019}.
 \\

This analysis suggests that the faster migration of planets located around 0.3 au—often referred to as the "Warm Jupiter desert"- could be explained by their shorter migration timescales. Meanwhile, the slower migration of planets located between 3 and 10 au, coupled with their higher occurrence rates, highlights the interplay between migration dynamics and observed exoplanet demographics. Moreover, we note that the migration timescales for planets with $a>1$~au around M dwarfs are 2 to 3 times longer than those for planets orbiting FGK stars. This finding provides new insight into giant planet formation around low-mass stars, as slower migration timescales are crucial for their formation. For instance, the Bern planet formation model \citep{Burn2021} requires reduced migration rates around low-mass stars to enable giant planet growth.

We also note that close-in planets with semi-major axis $a<0.1$~au, commonly referred to as "Hot Jupiters", are associated with low occurrence rates and very short migration timescales. The outward migration pathway identified in this work for planet-to-star mass ratios $q>0.002$ effectively rules out the formation of such planets via inward migration.
Therefore, their formation is more likely to result from high-eccentricity migration—triggered by planet-planet scattering followed by tidal circularization \citep[e.g.,][]{RasioFord1996, Ford2008, Nagasawa2008}. Alternatively, in situ formation has been proposed by other authors \citep[e.g.,][]{Boley2016, Bat2016}. However, this scenario remains challenging, as it would require planet formation to occur late in the disk’s evolution, when outward migration is expected to be negligible.\\

These insights provide a deeper understanding of the distribution of giant planets and the mechanisms shaping their final orbital configurations.


\section{Discussion}
\label{sec:discussion}

\begin{figure*}
    \centering
    \includegraphics[width=0.33\linewidth]{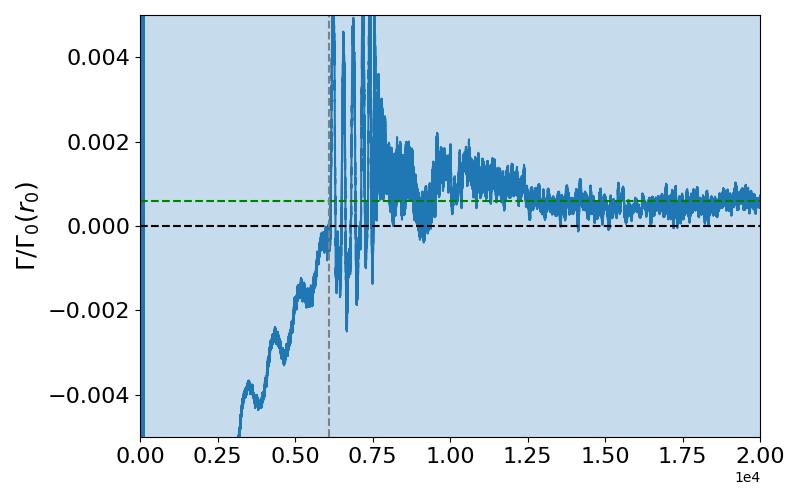}
    \includegraphics[width=0.33\linewidth]{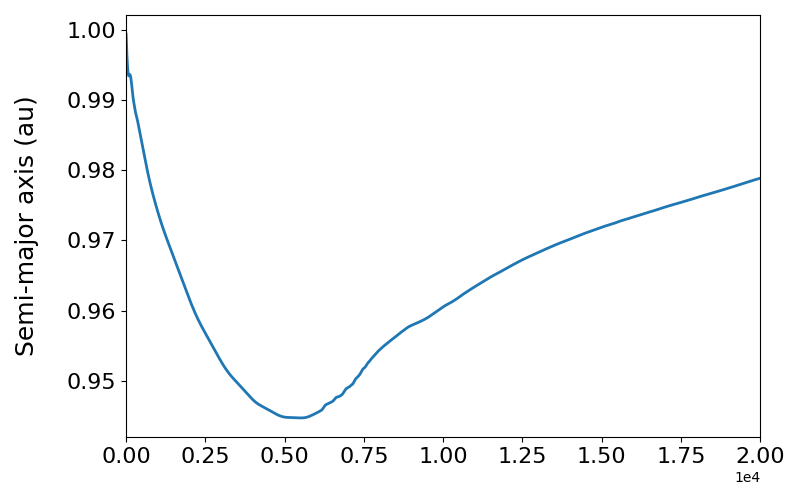}
    \includegraphics[width=0.33\linewidth]{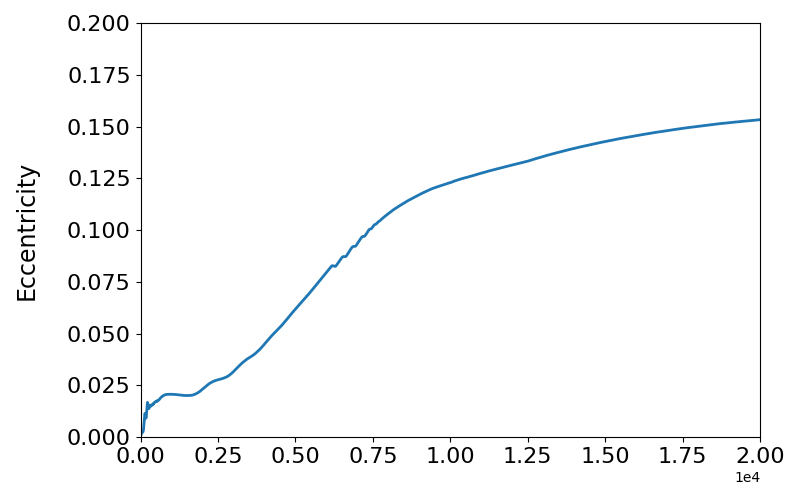}\\
    \includegraphics[width=0.33\linewidth]{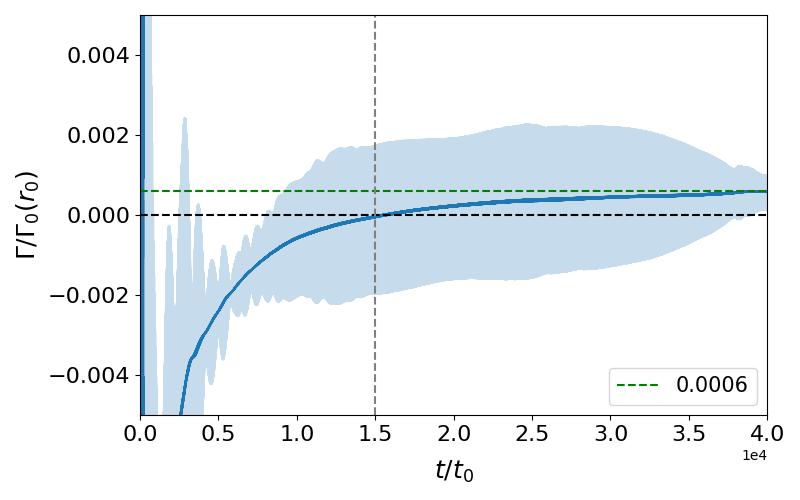}
    \includegraphics[width=0.33\linewidth]{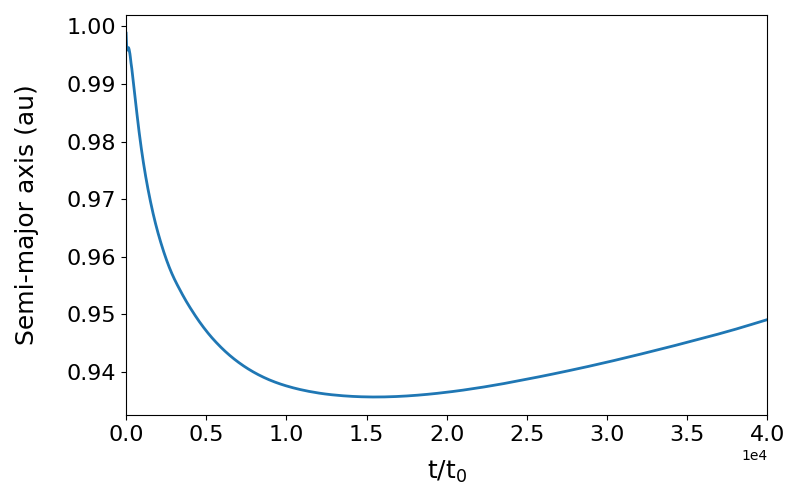}
    \includegraphics[width=0.33\linewidth]{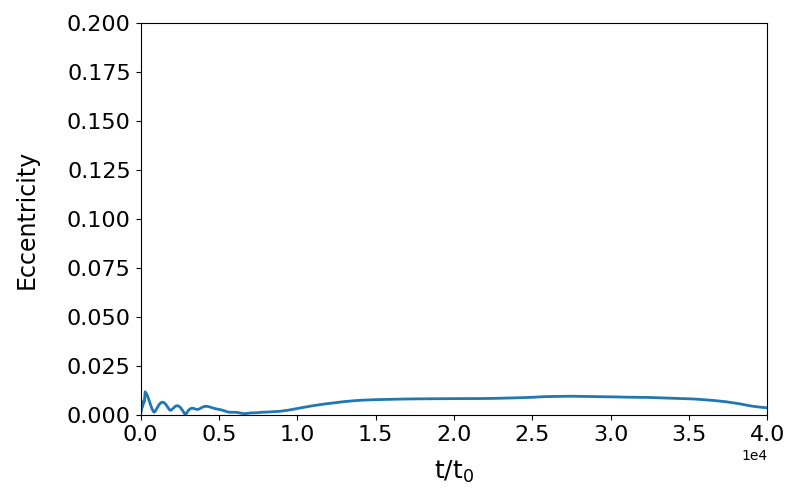}
    
    \caption{Comparison of the normalized torques (right panels), evolution of the semi-major axis (middle panels) and evolution of the eccentricity (left panels) of the planet from the case of study shown in Figure \ref{fig:standard-torques}, for two different resolutions: the one used in K18 (top) and the one used in this work (bottom).}
    \label{fig:HR-sim}
\end{figure*}

Our aim was to extend the analysis conducted by K18 to investigate the migration paths of single giant planets embedded in low-viscosity protoplanetary disks. However, due to the computational cost of such simulations, we chose to lower the resolution compared to the one used by K18. Prior to making this adjustment, we replicated several of K18's simulations to ensure that the normalized torques obtained with the reduced resolution were consistent with their results. As an example, in Figure \ref{fig:HR-sim}, we show the normalized torques along with the corresponding fitting function, as well as the evolution of the semi-major axis and eccentricity for a planet with $q=3 \times 10^{-3}$, using the same scenario as in Figure \ref{fig:standard-torques}. Two different resolutions are considered: the one used in this work (see Section \ref{sec:Method}) and the one proposed by K18, which is four times higher. The results confirm that in both cases, the same value of the normalized torque is achieved. With higher resolution, the planet reaches this value more quickly in its orbital evolution. In both cases, once the torque changes sign and becomes positive, the semi-major axis evolution shows that the planet begins migrating outward, and eccentricity is $e<0.2$. Furthermore, we note that the normalized torque exhibits oscillations over time, which are linked to the planet’s epicyclic motion. In particular, in the high-resolution case, as the planet reaches a higher eccentricity, the epicyclic motion has a larger amplitude, which leads to torque oscillations of greater amplitude. \\

In FARGO3D, the gas is not self-gravitating and orbits only under the influence of the star’s potential, while an embedded planet experiences a combined stellar and disk potential. This discrepancy shifts the Lindblad and corotation resonances, affecting the planet's migration velocity, as demonstrated by \cite{Baruteau2008}. A solution to this is to remove the azimuthally averaged density from the density of each zone before evaluating the forces, allowing the planet to feel only the star's gravity \citep[see e.g.,][]{Benitez-Llambay2016}. In order to compare with K18, we did not subtract the asymmetric component of the disc’s gravity from the force on the planet. However, to test the impact of this effect for high planet-star mass ratios, we activate the subtraction of this effect. Our results showed that this leads to smoother evolution of the torques and semi-major axis, and while the differences get larger with increased surface density, qualitatively the results do not change, as for the high mass ratios ($0.002<q<0.02$) planets still migrate outward at a comparable speed.\\

 We note that following K18, we did not include the disk’s indirect term in our main simulations. While it can be argued that it should be included on physical grounds, we found that it can also lead to artificial growth of eccentricity in the disk, resulting in inaccurate planetary migration behavior (see Appendix \ref{sec:ap}). 
Given its potential relevance \citep[e.g.,][]{Benitez-Llambay2016, Crida2025}, we performed test
runs to assess its impact on the planet’s dynamical evolution and and on the final torque estimates. In these tests, the disk's indirect term is applied consistently to both the gas and the planet. As a result, for planets with $e < 0.2$, we found no qualitative differences in the torque measurements during the quasi-steady-state phase, density maps, or migration direction—suggesting that the system behaves similarly whether or not the indirect term is included.
In a more extreme case with $q=0.01$ and $\Sigma_0=10^{-3}$, including the indirect term introduces a positive torque contribution. 
Still, migration stalls in both cases, and the planet’s orbit remains stable, indicating that long-term evolution is not significantly affected.
An intermediate case with $q=0.01$ and $\Sigma_0=10^{-4}$ shows lower eccentricities when the indirect term is included with values $e<0.25$. In summary, including the indirect term in the disk has only a modest impact on the overall migration behavior in the scenarios studied. Although it slightly modifies the early torque evolution and eccentricity damping, the long-term trends, such as migration direction and the final strength of the torques, remain broadly consistent, and within the scatter considered in our analysis (see further details in the Appendix \ref{sec:ap}).\\

The torques measured in the simulations arise from the gas perturbations inside and outside of the planet’s orbit. The boundary conditions of our simulations, particularly those at the inner edge of the disk, can influence the torques. For example, adjusting the inner boundary of the disk closer to the star (from 0.4r$_0$ to 0.2r$_0$) reduces the positive contribution from the inner disk and primarily impacts systems with smaller planet-star mass ratios. However, planets with the highest mass ratios ($0.01 < q < 0.02$) and correspondingly high $K$-values ($K > 3 \times 10^{6}$) exhibit increasing eccentricity ($e > 0.2$), despite all planets initially having $e = 0$. Nevertheless, the total torque and overall disk-planet dynamics remain qualitatively unchanged.
This underscores the need of realistic boundary conditions when modeling planet-disk interactions of massive planets. Since the evolution of gaps and the disk structure depends on viscous time scales, boundary conditions directly affect the dynamics \citep[see e.g ][]{Dempsey2020, 2024arXiv241000374B}.\\

The planets set in our simulations were allowed to migrate but not to accrete gas from the surrounded disk, following K18. However, \citet{Robert2018} showed that including gas accretion onto the planet—while simultaneously removing the accreted gas from the simulation—can facilitate gap opening and allow the system to reach a quasi-steady state earlier in the planet’s evolution, particularly in disks with higher viscosities. More recently, \cite{WuChen2025} investigated the effects of thermodynamics of a migrating an accreting giant planet of one Jupiter mass. They found that the total torque exerted onto the accreting planet directly depends on the $\beta$-cooling factor, which is associated with local isothermal conditions. This dependence can lead to a shift from inward to outward migration for certain parameter sets. Thus, it would be interesting to explore in future work the role of gas accretion for high planet-to-star mass ratios, particularly to assess whether gap opening and the onset of a quasi-steady state can occur earlier in the planetary evolution in low-viscosity disks. \\

Following \cite{Dempsey2021}, we observe that for the high planet-to-star mass ratios considered in this work, the gaps are significantly wider than the disk scale height. This suggests that the 2D treatment in our hydrodynamical simulations effectively captures the generation and transfer of torques. Consistent with their findings, we find that for Sun-like stars, super-Jupiter-mass planets may migrate outward to wider orbits. Moreover, extrapolating our results to M dwarfs suggests that planets more massive than 0.2 Jupiter masses could also end up in wider orbits.

\section{Summary and conclusions}
\label{sec:conclusions}

In this work, we performed hydrodynamical simulations using the \textsc{FARGO3D} code to analyze the migration speed and direction of planets with high planet-star mass ratios ($5 \times 10^{-4}< q < 2 \times 10^{-2}$, assuming a low-viscosity protoplanetary disk ($\alpha=10^{-4}$). We explored different sets of disk physical properties by varying the local gas surface density, the disk scale height, and the density slope index, which is also linked to the scale height profile. In all simulations, we assumed a single migrating planet that does not accrete gas from the disk.

We analyzed the resulting torques once they reached close to a steady state, as well as the planet’s orbital evolution in terms of semi-major axis and eccentricity. We classified the torque values in each scenario based on their magnitude and sign, associating them with inward or outward migration and linking them to the planet’s eccentricity. Additionally, we proposed analytical prescriptions to describe the migration of planets with high planet-star mass ratios that are in quasi-circular orbits and in low-viscosity disks. Furthermore, we compared the migration timescales of the studied planets with the occurrence rates of massive planets around stars of 1 M$_\odot$ and 0.1 M$_\odot$.\\
Our key findings are summarized below:\\
\begin{itemize}
 \item Planets associated to $q<0.0015$ experience inward migration. This represents planetary masses below 1.5 M$_{Jupiter}$ around Sun-like stars, and planets below 0.15 M$_{Jupiter}$ around M dwarfs.\\
 \item Planets with $q$ values between 0.0015 and 0.002 may experience a halt in migration, depending on the disk’s physical properties.\\
\item Planets with $q>0.002$ consistently experience outward migration, driven by a positive torque, as long as their eccentricity remains below $e < 0.2$, and $K>10^{4}$. Moreover, their migration speed depends on the local gas density.\\
\item For cases with $\Sigma_0 \leq 10^{-4}$ and $h_0 \leq 0.05$, planets with $q = 0.01$ and $q = 0.02$ develop eccentricities in the range $0.2 < e < 0.45$. Migration generally stalls once their eccentricities stabilize over time.\\
\item The migration timescales derived from the normalized torques in our simulations associated to planets in quasi-circular orbits, show a correlation with the occurrence rates of giant planets with $a<10$~au around FGK stars and also M dwarfs. Planets located closer to the star (around 0.3 au) exhibit shorter migration timescales (faster migration), where the giant planet occurrence rate is low, whereas planets farther away (3–10 au) have longer timescales (slower migration), where the giant planet occurrence rate is high.\\
\end{itemize}

Our study underscores the strong dependence of migration direction on the planet-to-star mass ratio, revealing a transition at $q \approx 0.002$ that separates inward from outward migration. These results enhance our understanding of the formation and evolution of super-Jupiter-like planets around Sun-like stars, as well as planets exceeding 0.2 Jupiter masses around very low-mass stars.
Given the challenges of detecting massive planets around low-mass stars, improving our theoretical understanding of their migration in low-viscosity disks is essential. This will provide insights into their formation and final orbits while guiding future observations and refining exoplanet population models.

\begin{acknowledgements}
This work was performed using the compute resources from
the Academic Leiden Interdisciplinary Cluster Environment (ALICE) provided
by Leiden University.
      M.~B.~S acknowledges travel support from FONDECYT project 11221206
      P.~B.~L. acknowledges  support from ANID, QUIMAL fund ASTRO21-0039 and FONDECYT project 1231205. The authors thank the referee for their valuable comments that helped to improve the quality of the article.
\end{acknowledgements}

%
%


\bibliographystyle{aa} 
\bibliography{biblioUp} 

\begin{thebibliography}{83}
\expandafter\ifx\csname natexlab\endcsname\relax\def\natexlab#1{#1}\fi

\bibitem[{{Bae} {et~al.}(2017){Bae}, {Zhu}, \& {Hartmann}}]{Bae2017}
{Bae}, J., {Zhu}, Z., \& {Hartmann}, L. 2017, \apj, 850, 201

\bibitem[{{Baruteau} {et~al.}(2014){Baruteau}, {Crida}, {Paardekooper},
  {Masset}, {Guilet}, {Bitsch}, {Nelson}, {Kley}, \&
  {Papaloizou}}]{Baruteau2014}
{Baruteau}, C., {Crida}, A., {Paardekooper}, S.~J., {et~al.} 2014, in
  Protostars and Planets VI, ed. H.~{Beuther}, R.~S. {Klessen}, C.~P.
  {Dullemond}, \& T.~{Henning}, 667--689

\bibitem[{{Baruteau} \& {Masset}(2008)}]{Baruteau2008}
{Baruteau}, C. \& {Masset}, F. 2008, \apj, 678, 483

\bibitem[{{Batygin} {et~al.}(2016){Batygin}, {Bodenheimer}, \&
  {Laughlin}}]{Bat2016}
{Batygin}, K., {Bodenheimer}, P.~H., \& {Laughlin}, G.~P. 2016, \apj, 829, 114

\bibitem[{{Ben{\'\i}tez-Llambay}(2024)}]{2024arXiv241000374B}
{Ben{\'\i}tez-Llambay}, P. 2024, arXiv e-prints, arXiv:2410.00374

\bibitem[{{Ben{\'\i}tez-Llambay} \& {Masset}(2016)}]{Pablo2016}
{Ben{\'\i}tez-Llambay}, P. \& {Masset}, F.~S. 2016, \apjs, 223, 11

\bibitem[{{Ben{\'\i}tez-Llambay} {et~al.}(2016){Ben{\'\i}tez-Llambay}, {Ramos},
  {Beaug{\'e}}, \& {Masset}}]{Benitez-Llambay2016}
{Ben{\'\i}tez-Llambay}, P., {Ramos}, X.~S., {Beaug{\'e}}, C., \& {Masset},
  F.~S. 2016, \apj, 826, 13

\bibitem[{{Bergsten} {et~al.}(2022){Bergsten}, {Pascucci}, {Mulders},
  {Fernandes}, \& {Koskinen}}]{Bergsten2022}
{Bergsten}, G.~J., {Pascucci}, I., {Mulders}, G.~D., {Fernandes}, R.~B., \&
  {Koskinen}, T.~T. 2022, \aj, 164, 190

\bibitem[{{Boley} {et~al.}(2016){Boley}, {Granados Contreras}, \&
  {Gladman}}]{Boley2016}
{Boley}, A.~C., {Granados Contreras}, A.~P., \& {Gladman}, B. 2016, \apjl, 817,
  L17

\bibitem[{{Bonfils} {et~al.}(2013){Bonfils}, {Delfosse}, {Udry}, {Forveille},
  {Mayor}, {Perrier}, {Bouchy}, {Gillon}, {Lovis}, {Pepe}, {Queloz}, {Santos},
  {S{\'e}gransan}, \& {Bertaux}}]{Bon2013}
{Bonfils}, X., {Delfosse}, X., {Udry}, S., {et~al.} 2013, \aap, 549, A109

\bibitem[{{Clanton} \& {Gaudi}(2014)}]{Clanton2014}
{Clanton}, C. \& {Gaudi}, B.~S. 2014, \apj, 791, 91

\bibitem[{{Cresswell} {et~al.}(2007){Cresswell}, {Dirksen}, {Kley}, \&
  {Nelson}}]{Cresswell2007}
{Cresswell}, P., {Dirksen}, G., {Kley}, W., \& {Nelson}, R.~P. 2007, \aap, 473,
  329

\bibitem[{{Crida} {et~al.}(2025){Crida}, {Baruteau}, {Griveaud}, {Lega},
  {Masset}, {B{\'e}thune}, {Fang}, {Gonzalez}, {M{\'e}heut}, {Morbidelli},
  {Gerosa}, {Kloster}, {Marques}, {Miniussi}, {Minker}, {Pichierri}, \&
  {Segretain}}]{Crida2025}
{Crida}, A., {Baruteau}, C., {Griveaud}, P., {et~al.} 2025, The Open Journal of
  Astrophysics, 8, 84

\bibitem[{{Crida} {et~al.}(2009){Crida}, {Baruteau}, {Kley}, \&
  {Masset}}]{Crida2009}
{Crida}, A., {Baruteau}, C., {Kley}, W., \& {Masset}, F. 2009, \aap, 502, 679

\bibitem[{{Crida} {et~al.}(2006){Crida}, {Morbidelli}, \& {Masset}}]{Crida2006}
{Crida}, A., {Morbidelli}, A., \& {Masset}, F. 2006, \icarus, 181, 587

\bibitem[{{Crida} {et~al.}(2007){Crida}, {Morbidelli}, \& {Masset}}]{Crida2007}
{Crida}, A., {Morbidelli}, A., \& {Masset}, F. 2007, \aap, 461, 1173

\bibitem[{{de Val-Borro} {et~al.}(2006){de Val-Borro}, {Edgar}, {Artymowicz},
  {Ciecielag}, {Cresswell}, {D'Angelo}, {Delgado-Donate}, {Dirksen}, {Fromang},
  {Gawryszczak}, {Klahr}, {Kley}, {Lyra}, {Masset}, {Mellema}, {Nelson},
  {Paardekooper}, {Peplinski}, {Pierens}, {Plewa}, {Rice}, {Sch{\"a}fer}, \&
  {Speith}}]{Valborro2006}
{de Val-Borro}, M., {Edgar}, R.~G., {Artymowicz}, P., {et~al.} 2006, \mnras,
  370, 529

\bibitem[{{Dempsey} {et~al.}(2020){Dempsey}, {Lee}, \&
  {Lithwick}}]{Dempsey2020}
{Dempsey}, A.~M., {Lee}, W.-K., \& {Lithwick}, Y. 2020, \apj, 891, 108

\bibitem[{{Dempsey} {et~al.}(2021){Dempsey}, {Mu{\~n}oz}, \&
  {Lithwick}}]{Dempsey2021}
{Dempsey}, A.~M., {Mu{\~n}oz}, D.~J., \& {Lithwick}, Y. 2021, \apjl, 918, L36

\bibitem[{{Dong} {et~al.}(2017){Dong}, {Li}, {Chiang}, \& {Li}}]{Dong2017}
{Dong}, R., {Li}, S., {Chiang}, E., \& {Li}, H. 2017, \apj, 843, 127

\bibitem[{{Dong} {et~al.}(2018){Dong}, {Li}, {Chiang}, \& {Li}}]{Dong2018}
{Dong}, R., {Li}, S., {Chiang}, E., \& {Li}, H. 2018, \apj, 866, 110

\bibitem[{{Duffell} {et~al.}(2014){Duffell}, {Haiman}, {MacFadyen}, {D'Orazio},
  \& {Farris}}]{Duffell2014}
{Duffell}, P.~C., {Haiman}, Z., {MacFadyen}, A.~I., {D'Orazio}, D.~J., \&
  {Farris}, B.~D. 2014, \apjl, 792, L10

\bibitem[{{Duffell} \& {MacFadyen}(2013)}]{Duffell2013}
{Duffell}, P.~C. \& {MacFadyen}, A.~I. 2013, \apj, 769, 41

\bibitem[{{D{\"u}rmann} \& {Kley}(2015{\natexlab{a}})}]{DKley2015}
{D{\"u}rmann}, C. \& {Kley}, W. 2015{\natexlab{a}}, \aap, 574, A52

\bibitem[{{D{\"u}rmann} \& {Kley}(2015{\natexlab{b}})}]{Durman2015}
{D{\"u}rmann}, C. \& {Kley}, W. 2015{\natexlab{b}}, \aap, 574, A52

\bibitem[{{Emsenhuber} {et~al.}(2021){Emsenhuber}, {Mordasini}, {Burn},
  {Alibert}, {Benz}, \& {Asphaug}}]{Burn2021}
{Emsenhuber}, A., {Mordasini}, C., {Burn}, R., {et~al.} 2021, \aap, 656, A69

\bibitem[{{Fernandes} {et~al.}(2019){Fernandes}, {Mulders}, {Pascucci},
  {Mordasini}, \& {Emsenhuber}}]{Fernandes2019}
{Fernandes}, R.~B., {Mulders}, G.~D., {Pascucci}, I., {Mordasini}, C., \&
  {Emsenhuber}, A. 2019, \apj, 874, 81

\bibitem[{{Flaherty} {et~al.}(2020){Flaherty}, {Hughes}, {Simon}, {Qi}, {Bai},
  {Bulatek}, {Andrews}, {Wilner}, \& {K{\'o}sp{\'a}l}}]{F2020}
{Flaherty}, K., {Hughes}, A.~M., {Simon}, J.~B., {et~al.} 2020, \apj, 895, 109

\bibitem[{{Flaherty} {et~al.}(2018){Flaherty}, {Hughes}, {Teague}, {Simon},
  {Andrews}, \& {Wilner}}]{F2018}
{Flaherty}, K.~M., {Hughes}, A.~M., {Teague}, R., {et~al.} 2018, \apj, 856, 117

\bibitem[{{Ford} \& {Rasio}(2008)}]{Ford2008}
{Ford}, E.~B. \& {Rasio}, F.~A. 2008, \apj, 686, 621

\bibitem[{{Fulton} {et~al.}(2021){Fulton}, {Rosenthal}, {Hirsch}, {Isaacson},
  {Howard}, {Dedrick}, {Sherstyuk}, {Blunt}, {Petigura}, {Knutson}, {Behmard},
  {Chontos}, {Crepp}, {Crossfield}, {Dalba}, {Fischer}, {Henry}, {Kane},
  {Kosiarek}, {Marcy}, {Rubenzahl}, {Weiss}, \& {Wright}}]{Fulton2021}
{Fulton}, B.~J., {Rosenthal}, L.~J., {Hirsch}, L.~A., {et~al.} 2021, \apjs,
  255, 14

\bibitem[{{Goldreich} \& {Sari}(2003)}]{Goldreich2003}
{Goldreich}, P. \& {Sari}, R. 2003, \apj, 585, 1024

\bibitem[{{Goodchild} \& {Ogilvie}(2006)}]{Goodchild2006}
{Goodchild}, S. \& {Ogilvie}, G. 2006, \mnras, 368, 1123

\bibitem[{{Griveaud} {et~al.}(2023){Griveaud}, {Crida}, \&
  {Lega}}]{Griveaud2023}
{Griveaud}, P., {Crida}, A., \& {Lega}, E. 2023, \aap, 672, A190

\bibitem[{{Guerra-Alvarado} {et~al.}(2025){Guerra-Alvarado}, {van der Marel},
  {Williams}, {Pinilla}, {Mulders}, {Lambrechts}, \& {Sanchez}}]{Osmar2025}
{Guerra-Alvarado}, O.~M., {van der Marel}, N., {Williams}, J.~P., {et~al.}
  2025, arXiv e-prints, arXiv:2503.19504

\bibitem[{{Hallam} \& {Paardekooper}(2018)}]{Hallam2018}
{Hallam}, P.~D. \& {Paardekooper}, S.~J. 2018, \mnras, 481, 1667

\bibitem[{{Ida} {et~al.}(2016){Ida}, {Guillot}, \& {Morbidelli}}]{Ida2016}
{Ida}, S., {Guillot}, T., \& {Morbidelli}, A. 2016, \aap, 591, A72

\bibitem[{{Ivanov} {et~al.}(1999){Ivanov}, {Papaloizou}, \&
  {Polnarev}}]{Ivanov1999}
{Ivanov}, P.~B., {Papaloizou}, J.~C.~B., \& {Polnarev}, A.~G. 1999, \mnras,
  307, 79

\bibitem[{{Kanagawa} {et~al.}(2017){Kanagawa}, {Tanaka}, {Muto}, \&
  {Tanigawa}}]{Kanagawa2017}
{Kanagawa}, K.~D., {Tanaka}, H., {Muto}, T., \& {Tanigawa}, T. 2017, \pasj, 69,
  97

\bibitem[{{Kanagawa} {et~al.}(2015){Kanagawa}, {Tanaka}, {Muto}, {Tanigawa}, \&
  {Takeuchi}}]{Kanagawa2015}
{Kanagawa}, K.~D., {Tanaka}, H., {Muto}, T., {Tanigawa}, T., \& {Takeuchi}, T.
  2015, \mnras, 448, 994

\bibitem[{{Kanagawa} {et~al.}(2018){Kanagawa}, {Tanaka}, \&
  {Szuszkiewicz}}]{Kanagawa2018}
{Kanagawa}, K.~D., {Tanaka}, H., \& {Szuszkiewicz}, E. 2018, \apj, 861, 140

\bibitem[{{Kimmig} {et~al.}(2020){Kimmig}, {Dullemond}, \& {Kley}}]{Kimmig2020}
{Kimmig}, C.~N., {Dullemond}, C.~P., \& {Kley}, W. 2020, \aap, 633, A4

\bibitem[{{Kley} \& {Dirksen}(2006)}]{Kley2006}
{Kley}, W. \& {Dirksen}, G. 2006, \aap, 447, 369

\bibitem[{{Laughlin} {et~al.}(2004){Laughlin}, {Bodenheimer}, \&
  {Adams}}]{Lau2004}
{Laughlin}, G., {Bodenheimer}, P., \& {Adams}, F.~C. 2004, \apjl, 612, L73

\bibitem[{{Lin} \& {Papaloizou}(1979)}]{Lin1979}
{Lin}, D.~N.~C. \& {Papaloizou}, J. 1979, \mnras, 186, 799

\bibitem[{{Lodato} {et~al.}(2019){Lodato}, {Dipierro}, {Ragusa}, {Long},
  {Herczeg}, {Pascucci}, {Pinilla}, {Manara}, {Tazzari}, {Liu}, {Mulders},
  {Harsono}, {Boehler}, {M{\'e}nard}, {Johnstone}, {Salyk}, {van der Plas},
  {Cabrit}, {Edwards}, {Fischer}, {Hendler}, {Nisini}, {Rigliaco}, {Avenhaus},
  {Banzatti}, \& {Gully-Santiago}}]{Lodato2019}
{Lodato}, G., {Dipierro}, G., {Ragusa}, E., {et~al.} 2019, \mnras, 486, 453

\bibitem[{{Lubow}(1991)}]{Lubow1991}
{Lubow}, S.~H. 1991, \apj, 381, 259

\bibitem[{{Masset}(2000)}]{Masset2000}
{Masset}, F. 2000, \aaps, 141, 165

\bibitem[{{Masset} \& {Snellgrove}(2001)}]{Masset2001}
{Masset}, F. \& {Snellgrove}, M. 2001, \mnras, 320, L55

\bibitem[{{McNally} {et~al.}(2019){McNally}, {Nelson}, {Paardekooper}, \&
  {Ben{\'\i}tez-Llambay}}]{McNally2019}
{McNally}, C.~P., {Nelson}, R.~P., {Paardekooper}, S.-J., \&
  {Ben{\'\i}tez-Llambay}, P. 2019, \mnras, 484, 728

\bibitem[{{Meyer} {et~al.}(2018){Meyer}, {Amara}, {Reggiani}, \&
  {Quanz}}]{Meyer2018}
{Meyer}, M.~R., {Amara}, A., {Reggiani}, M., \& {Quanz}, S.~P. 2018, \aap, 612,
  L3

\bibitem[{{Morbidelli} \& {Raymond}(2016)}]{Morbidelli2016}
{Morbidelli}, A. \& {Raymond}, S.~N. 2016, Journal of Geophysical Research
  (Planets), 121, 1962

\bibitem[{{Nagasawa} {et~al.}(2008){Nagasawa}, {Ida}, \&
  {Bessho}}]{Nagasawa2008}
{Nagasawa}, M., {Ida}, S., \& {Bessho}, T. 2008, \apj, 678, 498

\bibitem[{{Nelson} {et~al.}(2000){Nelson}, {Papaloizou}, {Masset}, \&
  {Kley}}]{Nelson2000}
{Nelson}, R.~P., {Papaloizou}, J. C.~B., {Masset}, F., \& {Kley}, W. 2000,
  \mnras, 318, 18

\bibitem[{{Nielsen} {et~al.}(2019){Nielsen}, {De Rosa}, {Macintosh}, {Wang},
  {Ruffio}, {Chiang}, {Marley}, {Saumon}, {Savransky}, {Ammons}, {Bailey},
  {Barman}, {Blain}, {Bulger}, {Burrows}, {Chilcote}, {Cotten}, {Czekala},
  {Doyon}, {Duch{\^e}ne}, {Esposito}, {Fabrycky}, {Fitzgerald}, {Follette},
  {Fortney}, {Gerard}, {Goodsell}, {Graham}, {Greenbaum}, {Hibon}, {Hinkley},
  {Hirsch}, {Hom}, {Hung}, {Dawson}, {Ingraham}, {Kalas}, {Konopacky},
  {Larkin}, {Lee}, {Lin}, {Maire}, {Marchis}, {Marois}, {Metchev},
  {Millar-Blanchaer}, {Morzinski}, {Oppenheimer}, {Palmer}, {Patience},
  {Perrin}, {Poyneer}, {Pueyo}, {Rafikov}, {Rajan}, {Rameau}, {Rantakyr{\"o}},
  {Ren}, {Schneider}, {Sivaramakrishnan}, {Song}, {Soummer}, {Tallis},
  {Thomas}, {Ward-Duong}, \& {Wolff}}]{Nielsen2019}
{Nielsen}, E.~L., {De Rosa}, R.~J., {Macintosh}, B., {et~al.} 2019, \aj, 158,
  13

\bibitem[{{Ogilvie}(2001)}]{Ogilvie2001}
{Ogilvie}, G.~I. 2001, \mnras, 325, 231

\bibitem[{{Paardekooper} {et~al.}(2010){Paardekooper}, {Baruteau}, {Crida}, \&
  {Kley}}]{Paardekooper2010}
{Paardekooper}, S.~J., {Baruteau}, C., {Crida}, A., \& {Kley}, W. 2010, \mnras,
  401, 1950

\bibitem[{{Paardekooper} {et~al.}(2011){Paardekooper}, {Baruteau}, \&
  {Kley}}]{Paardekooer2011}
{Paardekooper}, S.~J., {Baruteau}, C., \& {Kley}, W. 2011, \mnras, 410, 293

\bibitem[{{Pan} {et~al.}(2024){Pan}, {Liu}, {Johansen}, {Ogihara}, {Wang},
  {Ji}, {Wang}, {Feng}, \& {Ribas}}]{Pan2024}
{Pan}, M., {Liu}, B., {Johansen}, A., {et~al.} 2024, \aap, 682, A89

\bibitem[{{Papaloizou} {et~al.}(2001){Papaloizou}, {Nelson}, \&
  {Masset}}]{Papa2001}
{Papaloizou}, J.~C.~B., {Nelson}, R.~P., \& {Masset}, F. 2001, \aap, 366, 263

\bibitem[{{Pizzati} {et~al.}(2023){Pizzati}, {Rosotti}, \&
  {Tabone}}]{Pizzati2023}
{Pizzati}, E., {Rosotti}, G.~P., \& {Tabone}, B. 2023, \mnras, 524, 3184

\bibitem[{{Rasio} \& {Ford}(1996)}]{RasioFord1996}
{Rasio}, F.~A. \& {Ford}, E.~B. 1996, Science, 274, 954

\bibitem[{{Robert} {et~al.}(2018){Robert}, {Crida}, {Lega}, {M{\'e}heut}, \&
  {Morbidelli}}]{Robert2018}
{Robert}, C.~M.~T., {Crida}, A., {Lega}, E., {M{\'e}heut}, H., \& {Morbidelli},
  A. 2018, \aap, 617, A98

\bibitem[{{Rosotti}(2023)}]{Rosotti2023}
{Rosotti}, G.~P. 2023, \nar, 96, 101674

\bibitem[{{Sabotta} {et~al.}(2021){Sabotta}, {Schlecker}, {Chaturvedi},
  {Guenther}, {Mu{\~n}oz Rodr{\'\i}guez}, {Mu{\~n}oz S{\'a}nchez}, {Caballero},
  {Shan}, {Reffert}, {Ribas}, {Reiners}, {Hatzes}, {Amado}, {Klahr}, {Morales},
  {Quirrenbach}, {Henning}, {Dreizler}, {Pall{\'e}}, {Perger}, {Azzaro},
  {Jeffers}, {Kaminski}, {K{\"u}rster}, {Lafarga}, {Montes}, {Passegger}, \&
  {Zechmeister}}]{Sabotta2021}
{Sabotta}, S., {Schlecker}, M., {Chaturvedi}, P., {et~al.} 2021, \aap, 653,
  A114

\bibitem[{{Sanchez} {et~al.}(2024){Sanchez}, {van der Marel}, {Lambrechts},
  {Mulders}, \& {Guerra-Alvarado}}]{Sanchez2024}
{Sanchez}, M., {van der Marel}, N., {Lambrechts}, M., {Mulders}, G.~D., \&
  {Guerra-Alvarado}, O.~M. 2024, \aap, 689, A236

\bibitem[{{Savitzky} \& {Golay}(1964)}]{Savitzky1964}
{Savitzky}, A. \& {Golay}, M.~J.~E. 1964, Analytical Chemistry, 36, 1627

\bibitem[{{Scardoni} {et~al.}(2022){Scardoni}, {Clarke}, {Rosotti}, {Booth},
  {Alexander}, \& {Ragusa}}]{Scardoni2022}
{Scardoni}, C.~E., {Clarke}, C.~J., {Rosotti}, G.~P., {et~al.} 2022, \mnras,
  514, 5478

\bibitem[{{Shakura} \& {Sunyaev}(1973)}]{Shakura1973}
{Shakura}, N.~I. \& {Sunyaev}, R.~A. 1973, \aap, 500, 33

\bibitem[{{Stef{\'a}nsson} {et~al.}(2023){Stef{\'a}nsson}, {Mahadevan},
  {Miguel}, {Robertson}, {Delamer}, {Kanodia}, {Ca{\~n}as}, {Winn}, {Ninan},
  {Terrien}, {Holcomb}, {Ford}, {Zawadzki}, {Bowler}, {Bender}, {Cochran},
  {Diddams}, {Endl}, {Fredrick}, {Halverson}, {Hearty}, {Hill}, {Lin},
  {Metcalf}, {Monson}, {Ramsey}, {Roy}, {Schwab}, {Wright}, \&
  {Zeimann}}]{Stef2023}
{Stef{\'a}nsson}, G., {Mahadevan}, S., {Miguel}, Y., {et~al.} 2023, Science,
  382, 1031

\bibitem[{{Syer} \& {Clarke}(1995)}]{Syer1995}
{Syer}, D. \& {Clarke}, C.~J. 1995, \mnras, 277, 758

\bibitem[{{Tanaka} {et~al.}(2002){Tanaka}, {Takeuchi}, \& {Ward}}]{Tanaka2002}
{Tanaka}, H., {Takeuchi}, T., \& {Ward}, W.~R. 2002, \apj, 565, 1257

\bibitem[{{Toci} {et~al.}(2021){Toci}, {Rosotti}, {Lodato}, {Testi}, \&
  {Trapman}}]{Toci2021}
{Toci}, C., {Rosotti}, G., {Lodato}, G., {Testi}, L., \& {Trapman}, L. 2021,
  \mnras, 507, 818

\bibitem[{{van der Marel}(2023)}]{Nienke2023}
{van der Marel}, N. 2023, European Physical Journal Plus, 138, 225

\bibitem[{{Vigan} {et~al.}(2017){Vigan}, {Bonavita}, {Biller}, {Forgan},
  {Rice}, {Chauvin}, {Desidera}, {Meunier}, {Delorme}, {Schlieder}, {Bonnefoy},
  {Carson}, {Covino}, {Hagelberg}, {Henning}, {Janson}, {Lagrange}, {Quanz},
  {Zurlo}, {Beuzit}, {Boccaletti}, {Buenzli}, {Feldt}, {Girard}, {Gratton},
  {Kasper}, {Le Coroller}, {Mesa}, {Messina}, {Meyer}, {Montagnier},
  {Mordasini}, {Mouillet}, {Moutou}, {Reggiani}, {Segransan}, \&
  {Thalmann}}]{Vigan2017}
{Vigan}, A., {Bonavita}, M., {Biller}, B., {et~al.} 2017, \aap, 603, A3

\bibitem[{{Villenave} {et~al.}(2020){Villenave}, {M{\'e}nard}, {Dent},
  {Duch{\^e}ne}, {Stapelfeldt}, {Benisty}, {Boehler}, {van der Plas}, {Pinte},
  {Telkamp}, {Wolff}, {Flores}, {Lesur}, {Louvet}, {Riols}, {Dougados},
  {Williams}, \& {Padgett}}]{Villaneva2020}
{Villenave}, M., {M{\'e}nard}, F., {Dent}, W.~R.~F., {et~al.} 2020, \aap, 642,
  A164

\bibitem[{{Villenave} {et~al.}(2022){Villenave}, {Stapelfeldt}, {Duch{\^e}ne},
  {M{\'e}nard}, {Lambrechts}, {Sierra}, {Flores}, {Dent}, {Wolff}, {Ribas},
  {Benisty}, {Cuello}, \& {Pinte}}]{Villaneva2022}
{Villenave}, M., {Stapelfeldt}, K.~R., {Duch{\^e}ne}, G., {et~al.} 2022, \apj,
  930, 11

\bibitem[{{Ward}(1997)}]{Ward1997}
{Ward}, W.~R. 1997, \icarus, 126, 261

\bibitem[{{Weber} {et~al.}(2019){Weber}, {P{\'e}rez}, {Ben{\'\i}tez-Llambay},
  {Gressel}, {Casassus}, \& {Krapp}}]{Weber2019}
{Weber}, P., {P{\'e}rez}, S., {Ben{\'\i}tez-Llambay}, P., {et~al.} 2019, \apj,
  884, 178

\bibitem[{{Wu} \& {Chen}(2025)}]{WuChen2025}
{Wu}, Y. \& {Chen}, Y.-X. 2025, \mnras, 536, L13

\bibitem[{{Ziampras} {et~al.}(2024{\natexlab{a}}){Ziampras}, {Nelson}, \&
  {Paardekooper}}]{Ziampras2024b}
{Ziampras}, A., {Nelson}, R.~P., \& {Paardekooper}, S.-J. 2024{\natexlab{a}},
  \mnras, 532, 351

\bibitem[{{Ziampras} {et~al.}(2024{\natexlab{b}}){Ziampras}, {Nelson}, \&
  {Paardekooper}}]{Ziampras2024a}
{Ziampras}, A., {Nelson}, R.~P., \& {Paardekooper}, S.-J. 2024{\natexlab{b}},
  \mnras, 528, 6130

\bibitem[{{Ziampras} {et~al.}(2025){Ziampras}, {Nelson}, \&
  {Paardekooper}}]{Ziampras2025}
{Ziampras}, A., {Nelson}, R.~P., \& {Paardekooper}, S.-J. 2025, arXiv e-prints,
  arXiv:2502.18564

\end{thebibliography}

\begin{appendix}

\section{Eccentricity evolution due to the indirect disk term}
\label{sec:ap}

We have worked in a stellocentric frame throughout. This frame is non-inertial, since the star is accelerated by gravity due to the planet, and, if the disk mass is considered, the disk. These accelerations appear as source terms in the momentum equations and are usually referred to as 'indirect terms'. Their forms are for example given in \cite{Nelson2000}.

While the indirect term due to the presence of a planet is very straightforwardly dealt with, the indirect term due to the disk poses interesting difficulties. For a start, most disk-planet interaction studies, including ours, neglect self-gravity of the disk. This would mean the disk technically has no mass, and therefore there would be no indirect term. However, the disk is can only cause the planet to migrate if it has mass, which leads to an inconsistency that needs to be corrected for by incorporating at least the axisymmetric component of de disk gravity if the planet is allowed to migrate \citep{Baruteau2008}. This inconsistency of the disk having mass in one sense (able to cause planet migration) and no mass in another sense (no self-gravity) also leads to the question of whether the indirect term due the disk should be included in simulations \citep{Crida2025}. Below, we show that it can be potentially problematic.

As in the rest of the paper, we consider a 2D disk, in cylindrical coordinates centered on the star:
\begin{align}
  \partial_t\Sigma + \nabla\cdot (\Sigma {\bf v})=& 0\\
  \partial_t {\bf v} + ({\bf v}\cdot\nabla){\bf v} =& -\frac{\nabla p}{\Sigma} - \nabla\Phi.
\end{align}
We take the equation of state to be strictly isothermal, but that is not important for the present discussion. The potential $\Phi $ has contributions from the central star, a direct part containing the contribution from the disk (if it is self-gravitating) and the planet (if present), and the indirect terms, again both due to disk and planet:
\begin{align}
  \Phi = -\frac{GM}{r} + \Phi_{\rm d, planet} +\Phi_{\rm i, planet}+ \Phi_{\rm d, disk} +\Phi_{\rm i, disk},
\end{align}
where the indirect potential due to the disk is given by
\begin{align}
  \Phi_{\rm i,disk} = Gr\int\int \Sigma(r',\varphi') \frac{\cos(\varphi-\varphi')}{r'}{\rm d} r' {\rm d}\varphi'.
\end{align}
The direct parts due to planet and disk and the indirect term due to the planet will not feature below. We write the 2D velocity vector as ${\bf v} = (u,v)^T$, and denote the angular velocity $\Omega = v/r$.

\subsection{Equilibrium state}

If we take the disk without a planet to be axisymmetric, with $u=0$, the equilibrium condition on the azimuthal velocity is simply
\begin{align}
  v^2 = r\partial_r\Phi + r\frac{\partial_r p}{\Sigma} = r^2\Omega_{\rm K}^2 + r\partial_r\Phi_{\rm d, disk} + \frac{r\partial_r p}{\Sigma},
\end{align}
where $\partial_r = \partial/\partial r$, $\Omega_{\rm K}$ is the Keplerian angular velocity and $\Sigma(r)$ is an arbitrary density profile. The penultimate term, denoting the axisymmetric component of the disk's gravity is the term causing inconsistency problems when considering migrating planets in non-self-gravitating disks, and needs to be compensated for \citep{Baruteau2008}. Defining a small parameter $\epsilon \sim H/r$, the last two terms are taken to be $O\left(\epsilon^2\right)$ and we can write $v = v_0 + v_2$ with
\begin{align}
  v_0 =& r\Omega_{\rm K},\\
  v_2 =& \frac{{\rm d}_r\Phi_{\rm d,disk} + {\rm d}_r p/\Sigma}{2\Omega_{\rm K}},
\end{align}
where ${\rm d}_r = {\rm d}/{\rm d}r$. \\

We note that the indirect term does not contribute for an axisymmetric disk.

\subsection{Linear perturbations}
We now add eccentric perturbations to the disk, of the form $\hat\Sigma(r) \exp(-{\rm i} \varphi - {\rm i}\omega t)$ and similar for other quantities, and we ignore terms that are quadratic in the perturbations:
\begin{align}
  -{\rm i} \omega \hat\xi + \frac{{\rm d}_r(r\hat u)-{\rm i}\hat v}{r} + \frac{{\rm d}_r\Sigma}{\Sigma} \hat u-{\rm i} \Omega\hat\xi=& 0\\
  -{\rm i}\omega \hat u -{\rm i} \Omega\hat u -2\Omega\hat v =& -c^2{\rm d}_r\hat\xi -{\rm d}_r\hat\Phi,\\
 -{\rm i}\omega\hat v + {\rm d}_rv\hat u -{\rm i}\Omega\hat v + \Omega\hat u=& \frac{{\rm i} c^2\hat\xi}{r} + \frac{{\rm i} \hat \Phi}{r},
\end{align}
with $c$ the sound speed, and $\hat\xi=\hat\Sigma/\Sigma$ the relative surface density perturbation. The potential perturbation contains terms due to the planet (direct and indirect) and due to the disk (direct, if we consider self-gravity, and indirect). The indirect term due to the disk is:
\begin{align}
  \hat \Phi_{\rm i,disk} = \pi Gr\int \frac{\hat\Sigma(r')}{r'}{\rm d} r'.
\end{align}
The exact form of the other parts of the potential are not important for our purposes.

\subsection{Eccentricity evolution}

We now assume time evolution to be slow, $\omega/\Omega_0 = O\left(\epsilon^2\right)$. When including a planet, this means we would work with an average potential and the corresponding secular evolution \citep[see e.g.][]{Goodchild2006}. We develop a series in $\epsilon \sim H/r \ll 1$, noting that the indirect term is of order $\epsilon^2$, comparable to the pressure terms. At lowest order in $\epsilon$, we find from the momentum equations that
\begin{align}
 -\Omega_0\hat u_0 +2{\rm i} \Omega_0\hat v_0=&0,\\
 -{\rm i}({\rm d}_rv_0+\Omega_0)\hat u_0 -\Omega_0\hat v_0=&0,
\end{align}
and making use of $v_0 \propto r^{-1/2}$ at lowest order, this implies that
$\hat u_0 =2{\rm i}\hat v_0 $, and we can write
\begin{align}
  \hat u_0 =& {\rm i} r \Omega_0 E(r),\\
  \hat v_0 =& \frac{1}{2} r\Omega_0 E(r),
\end{align}
with complex eccentricity $E=e\exp({\rm i} \varpi)$. Thus, at this level of approximation, fluid elements are on eccentric Keplerian orbits. Higher-order terms give an evolution equation for the eccentricity \citep{Ogilvie2001}, but this is not needed for this discussion.

From the continuity equation at lowest order we find that the surface density response in terms of the eccentricity is given by:
\begin{align}
  \hat\xi_0=r {\rm d}_r E+\frac{r{\rm d}_r\Sigma_0}{ \Sigma_0}  E
  = \frac{r}{\Sigma_0} {\rm d}_r\left(\Sigma_0E\right).
\end{align}
Interestingly, this makes the indirect potential
\begin{align}
  \hat \Phi_{\rm i,disk} = \pi Gr\int \frac{\Sigma_0}{r'}\hat\xi_0{\rm d} r'
  =\pi Gr\left[\Sigma_0E\right]_{r_{\rm in}}^{r_{\rm out}},
\end{align}
which means that the indirect term can only affect the eccentricity through the boundaries. If we consider a ring of material, where $\Sigma_0 \rightarrow 0$ for $r\rightarrow 0$ and $r\rightarrow \infty$, the indirect term will not contribute to the eccentricity evolution. If we consider only part of the disk, so that $\Sigma_0 \neq 0$ at the boundaries, the boundaries can act as a source of eccentricity, much like the 3:1 resonance considered in a binary context in \cite{Goodchild2006}. One would find exponential growth of eccentricity, driven by the boundaries. This seems artificial, as there is nothing special about the position of the boundaries, unlike the 3:1 resonance \citep{Lubow1991}. A sensible boundary condition could be to demand $E=0$, in which case the indirect term again does not contribute. This is however hard to enforce in practice, as it involves setting a hard boundary (i.e. reflective boundary conditions), which is undesirable in the presence of density waves excited by the planet and a viscous accretion flow onto the star. Moreover, any deviation from $E=0$ due to inexact boundary conditions will artificially add to the eccentricity evolution.

It is worth noting that this behaviour is independent of the inclusion of self-gravity. In a self-gravitating disk, there would be no question of whether to include the indirect term due to the disk (physics demands it), but the same spurious eccentricity growth could occur if not the whole disk is simulated. Of course, there is usually an implicit assumption in self-gravitating grid-based simulations that there is no mass outside the computational domain \citep[e.g.][]{Baruteau2008}, but the above discussion suggests that care needs to be taken at the boundaries.

Therefore, while including the indirect term due to the disk gravity might be justified on physical grounds, it can lead through artificial eccentricity growth through the boundaries of the domain.

\end{appendix}

\end{document}